\documentstyle[11pt,times,chicagor,smallsec,url,epsf]{article}

\newcommand{\<}{\langle}
\renewcommand{\>}{\rangle}

\newcommand{\G}{{\cal G}}
\newcommand{\J}{{\cal J}}
\newcommand{\F}{{\cal F}}
\renewcommand{\H}{H}

\newcommand{\I}{{\cal I}}

\renewcommand{\u}{{\sf u}}

\newcommand{\commentout}[1]{}
\newcommand{\A}{{A}}

\newcommand{\union}{\cup}
\newcommand{\inter}{\cap}

\commentout{
\newtheorem{THEOREM}{Theorem}[section]
\newenvironment{theorem}{\begin{THEOREM} \hspace{-.85em} {\bf :} }%
                        {\end{THEOREM}}

\newtheorem{DEFINITION}[THEOREM]{Definition}
\newenvironment{definition}{\begin{DEFINITION} \hspace{-.85em} {\bf :} \rm}%
                            {\end{DEFINITION}}

\newtheorem{COROLLARY}[THEOREM]{Corollary}
\newenvironment{corollary}{\begin{COROLLARY} \hspace{-.85em} {\bf :} \rm}%
                            {\end{COROLLARY}}

\newcommand{\thm}{\begin{theorem}}
\newcommand{\ethm}{\end{theorem}}
}
\newtheorem{theorem}{Theorem}[section]
\newtheorem{corollary}{Corollary}[section]
\newtheorem{lemma}{Lemma}[section]
\newtheorem{proposition}{Proposition}[section]
\newtheorem{definition}{Definition}[section]
\newtheorem{non-theorem}{Non-theorem}[section]

\newcommand{\thm}{\begin{theorem}}
\newcommand{\lem}{\begin{lemma}}
\newcommand{\pro}{\begin{proposition}}
\newcommand{\dfn}{\begin{definition} \rm}
\newcommand{\rem}{\begin{remark}}
\newcommand{\xam}{\begin{example}}
\newcommand{\cor}{\begin{corollary}}
\newcommand{\prf}{\noindent{\bf Proof:} }
\newcommand{\ethm}{\end{theorem}}
\newcommand{\elem}{\end{lemma}}
\newcommand{\epro}{\end{proposition}}
\newcommand{\edfn}{\bbox\end{definition}}
\newcommand{\erem}{\bbox\end{remark}}
\newcommand{\exam}{\bbox\end{example}}
\newcommand{\ecor}{\end{corollary}}
\newcommand{\eprf}{\bbox\vspace{0.1in}}
\newcommand{\beqn}{\begin{equation}}
\newcommand{\eeqn}{\end{equation}}

\newcommand{\bbox}{\vrule height7pt width4pt depth1pt}
\newcommand{\dom}{\mathit{dom}}

\newenvironment{oldthm}[1]{\par\noindent{\bf Theorem #1:} \em \noindent}{\par}
\newenvironment{oldlem}[1]{\par\noindent{\bf Lemma #1:} \em \noindent}{\par}
\newenvironment{oldcor}[1]{\par\noindent{\bf Corollary #1:} \em \noindent}{\par}
\newenvironment{oldpro}[1]{\par\noindent{\bf Proposition #1:} \em \noindent}{\par}
\newcommand{\othm}[1]{\begin{oldthm}{\ref{#1}}}
\newcommand{\eothm}{\end{oldthm} \medskip}
\newcommand{\olem}[1]{\begin{oldlem}{\ref{#1}}}
\newcommand{\eolem}{\end{oldlem} \medskip}
\newcommand{\ocor}[1]{\begin{oldcor}{\ref{#1}}}
\newcommand{\eocor}{\end{oldcor} \medskip}
\newcommand{\opro}[1]{\begin{oldpro}{\ref{#1}}}
\newcommand{\eopro}{\end{oldpro} \medskip}
\newcommand{\IR}{\mbox{$I\!\!R$}}

\newcommand{\shortv}{\commentout}
\newcommand{\fullv}[1]{#1}

\begin{document}

\title{Extensive Games with Possibly Unaware Players%
\thanks{
A preliminary version of this work was presented at AAMAS06
conference in Hakodate, Japan, in May of 2006. This work was
supported in part by NSF under grants CTC-0208535, ITR-0325453, and
IIS-0534064, by ONR under grants N00014-00-1-03-41 and
N00014-01-10-511, and by the DoD Multidisciplinary University
Research Initiative (MURI) program administered by the ONR under
grant N00014-01-1-0795. The second author was also supported in part
by a scholarship from the Brazilian Government through the Conselho
Nacional de Desenvolvimento Cient\'ifico e Tecnol\'ogico (CNPq). }}
%
%

%

\author{{\bf Joseph Y. Halpern}
\\ Computer Science Department \\ Cornell University,
U.S.A. \\  e-mail: halpern@cs.cornell.edu \\
 {\bf Leandro Chaves R\^ego}
\thanks{Most of this work was done while the author was at the School
of Electrical and Computer Engineering at Cornell University,
U.S.A.}
\\ Statistics Department \\
Federal University of Pernambuco, Brazil \\
e-mail: leandro@de.ufpe.br}

\date{}
\maketitle

\pagebreak

\begin{abstract}
Standard game theory assumes that the structure of the game is
common knowledge among players. We relax this assumption by
considering extensive games where agents may be unaware of the
complete structure of the game.  In particular, they may not be
aware of moves that they and other agents can make. We show how such
games can be represented; the key idea is to describe the game from
the point of view of every agent at every node of the game tree. We
provide a generalization of Nash equilibrium and show that every
game with awareness has a generalized Nash equilibrium. Finally, we
extend these results to games with \emph{awareness of unawareness},
where a player $i$ may be aware that a player $j$ can make moves
that $i$ is
not aware of%
\fullv{,
and to {\em subjective games}, where payers may have no common
knowledge regarding the actual game and their beliefs are
incompatible with a common prior}.
\end{abstract}

{\bf keywords:} Economic Theory, Foundations of Game Theory,
Awareness, Solution Concepts.


\section{Introduction}
\label{sec:intro}

Standard game theory models implicitly assume that all significant
aspects of the game (payoffs, moves available, etc.) are common
knowledge among the players. While such common knowledge may seem
unreasonable, there are well-known techniques going back to Harsanyi
\citeyear{Harsanyi} for transforming a game where some aspects are
not common knowledge to one where they are common knowledge.
All these techniques assume that players are at least \emph{aware}
of all possible moves in the game.
However, this is not always a reasonable assumption. For example,
sleazy companies assume that consumers are not aware that they can
lodge complaints if there are problems; in a war setting, having
technology that an enemy is unaware of (and thus being able to make
moves that the enemy is unaware of) can be critical; in financial
markets, some investors may not be aware of certain investment
strategies (complicated hedging strategies, for example, or
tax-avoidance strategies).

\commentout{
To understand the relevance of adding the possibility
of unawareness to the analysis of games, consider the game shown in
Figure~\ref{fig:game1}. One Nash equilibrium of this game has $A$
playing across$_A$ and $B$ playing down$_B$.  However, suppose that
$A$ is not aware that $B$ can play down$_B$.  In that case, if $A$
is rational, $A$ will play down$_A$.

\begin{figure}[ht]
\label{fig:game1} \centering \epsfxsize=9cm \epsffile{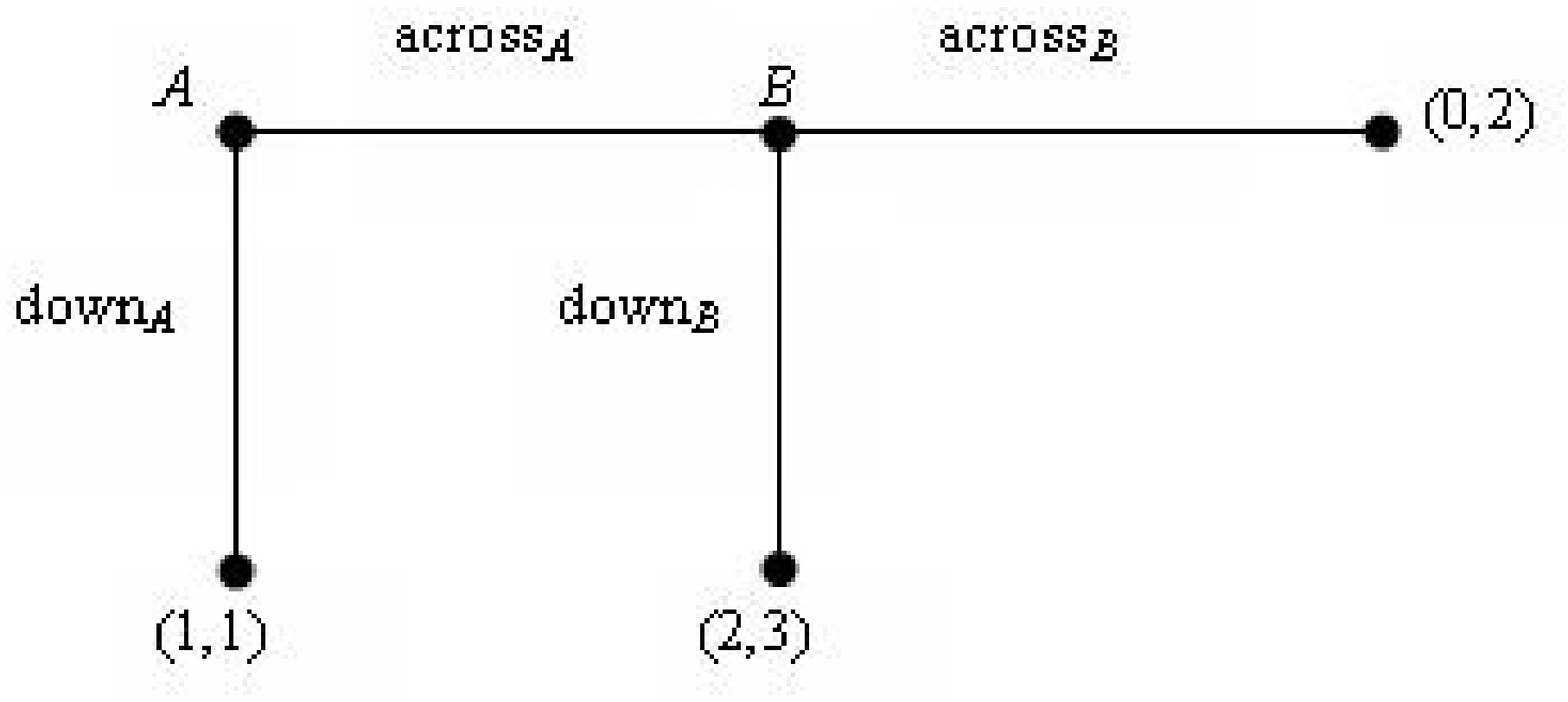}
\caption{A simple game.}
\end{figure}

A set of strategies is a Nash equilibrium if each agent's strategy
is a best response to the other agents' strategies, so each agent
$i$ would continue playing its strategy even if $i$ knew what
strategies the other agents were using.  This intuition does not
apply to (across$_A$,down$_B$).  Although $A$ would play across$_A$
if $A$ knew that $B$ were going to play down$_B$, $A$ cannot even
contemplate this possibility, let alone know it. Nash equilibrium
does not seem to be the appropriate solution concept here.
}
%

In a standard game, a set of strategies is a Nash equilibrium if each
agent's strategy is a best response to the other agents' strategies,
so each agent $i$ would continue playing its strategy even if $i$
knew what strategies the other agents were using. To understand the
relevance of adding the possibility of unawareness to the analysis
of games, consider the game shown in Figure~\ref{fig:game1}. One
Nash equilibrium of this game has $A$ playing across$_A$ and $B$
playing down$_B$.  However, suppose that $A$ is not aware that $B$
can play down$_B$.  In that case, if $A$ is rational, $A$ will play
down$_A$. Therefore, Nash equilibrium does not seem to be the
appropriate solution concept here. Although $A$ would play
across$_A$ if $A$ knew that $B$ were going to play down$_B$, $A$
cannot even contemplate this possibility, let alone know it.

\begin{figure}[ht]
\label{fig:game1} \centering \epsfxsize=12cm \epsffile{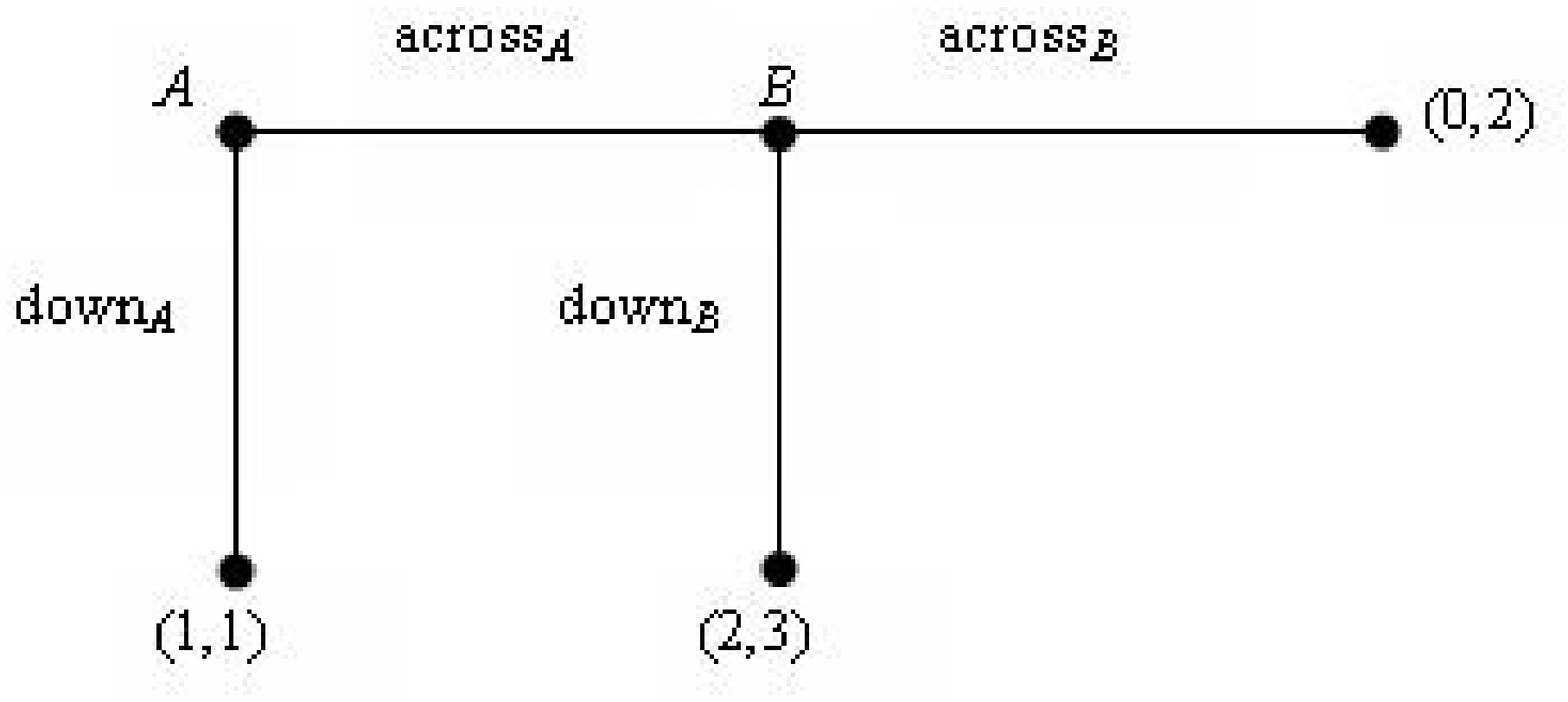}
\caption{A simple game.}
\end{figure}

Our goal is to find appropriate solution concepts for extensive
games with possibly unaware players, and more generally, to find
ways of representing multiagent systems where some agents may not be
aware of features of the system.  To do this, we must first find an
appropriate representation for such games. The first step in doing
so is to explicitly represent what players are aware of at each
node.  We do this by using what we call an \emph{augmented game}.
An augmented game describes how awareness changes over time.
For example, perhaps $A$ playing across$_A$ will result in $B$
becoming aware of the possibility of playing down$_B$. In financial
settings, one effect of players using certain investment strategies
is that other players become aware of the possibility of using that
strategy. Strategic thinking in such games must involve taking this
possibility into account.

We cannot in general represent what is going on using only one
augmented game. The standard representation of a game
implicitly assumes that (it is common knowledge that) the modeler and
the players all understand the game the same way.
This is no longer true once we allow for the possibility of
unawareness, since a player's description of the game can now
involve only those aspects of the game that he is aware of. Thus,
the full description of the game with awareness is given by a set of
augmented games, one for the modeler and one for each game that at
least one of the agents thinks might be the true game in some
situation.

Continuing with the game in Figure~\ref{fig:game1}, the augmented
game from the point of view of the type of $B$ that is unaware
of the possibility of playing down$_B$ would just include $A$'s
moves down$_A$ and across$_A$ and the move across$_B$.  In that
augmented game, player $A$ is also unaware of the move down$_B$. By
way of contrast, the augmented game from the point of view of the type
of $B$ that is
aware of down$_B$ would include the move down$_B$, but may also
allow for the possibility that $A$ is not aware that $B$ is aware of
this move.

The standard notion of Nash equilibrium consists of a collection of
strategies, one for each player.  Our generalization consists of a
collection of strategies, one for each pair $(i,\Gamma')$, where
$\Gamma'$ is a game that agent $i$ considers to be the true game in
some situation. Intuitively, the strategy for a player $i$ at
$\Gamma'$ is the strategy $i$ would play in situations where $i$
believes that the true game is $\Gamma'$. To understand why we may
need to consider different strategies consider, for example, the
game of Figure~\ref{fig:game1}. $B$ would play differently depending
on whether or not he was aware of down$_B$. Roughly speaking, a set
of strategies, one for each pair $(i,\Gamma')$, is a
\emph{generalized Nash equilibrium} if the strategy for
$(i,\Gamma')$ is a best response for player $i$ if the true game is
$\Gamma'$, given the strategies being used by the other players in
$\Gamma'$.

We argue that this notion of equilibrium correctly captures our
intuitions. We then show that every game with awareness has a
generalized Nash equilibrium by associating with a game with awareness a
standard game (where agents are aware of all moves) such
that there is a one-to-one correspondence between generalized Nash
equilibria of the game with awareness and Nash equilibria of the
standard game.

\commentout{ Many solution concepts besides Nash equilibrium have
been considered for standard games.  Our representation allows all
of them to be generalized to games with awareness in a
straightforward way.  We illustrate this
by considering one of the most frequently-used solution concepts,
\emph{sequential equilibrium} \cite{KW82}.
}

For ease of exposition, for most of the paper we focus on games
where agents are not aware of their lack of awareness.  That is, we
do not consider games where one player might be aware that there are
moves that another player (or even she herself) might be able to
make, although she is not aware of what they are. Such awareness of
unawareness can be quite relevant in practice. For example, in the
war setting described above, even if one side cannot conceive of a
new technology available to the enemy, they might believe that there
is some move available to the enemy without understanding what that
particular move is.  This, in turn, may encourage peace overtures.
To take another example, an agent might delay making a decision
because she considers it possible that she might learn about more
possible moves, even if she is not aware of what these moves are.

If we interpret ``lack of awareness'' as ``unable to compute''
(cf.~\cite{FH}), then awareness of unawareness becomes even more
significant.  Consider a chess game.  Although all players
understand in principle all the moves that can be made, they are
certainly not aware of all consequences of all moves.   A more
accurate representation of chess would model this computational
unawareness explicitly. We provide such a representation.

Roughly speaking, we capture the fact that player $i$ is aware that,
at a node $h$ in the game tree, there is a move that $j$ can make
she ($i$) is not aware by having $i$'s subjective representation of
the game include a ``virtual'' move for $j$ at node $h$.  Since $i$
might have only an incomplete understanding of what can happen after
this move,

$i$ simply describes what she believes will be the game after the
virtual move,
to the extent that she can.  For example, if she has no idea what will
happen after the virtual move, then she can describe her beliefs
regarding the payoffs of the game.
Thus, our representation can be viewed as a generalization of how chess
programs
analyze chess games.  They explore the game tree up to a certain
point, and then evaluate the board position at that point. We can
think of the payoffs following a virtual move by $j$ in $i$'s
subjective representation of a chess game as describing the
evaluation of the board from $i$'s point of view. This seems like a
much more reasonable representation of the game than the standard
complete game tree!

\fullv{
Our framework is flexible enough to deal with games where there is
lack of common knowledge about what is the game being played.
In particular, we can deal with lack of common knowledge regarding
the utilities, who moves next, the structure of other players'
information sets, and the probability of nature's moves (even in
cases where there is no common prior compatible with the players'
beliefs regarding nature).}

\commentout{Recently, Feinberg~\citeyear{Feinberg04,Feinberg05} also
studied games with awareness and defined a generalized notion of
Nash equilibrium in such games. Our approach is much simpler than
his, since his representation explicitly represents nested levels of
awareness.  That is, he explicitly describes whether player 1 is
aware of player 2's move, whether 2 is aware that 1 is aware of 2's
move, and so on. Moreover, our approach lends itself more naturally
to dealing with awareness of unawareness.}

Recently, Feinberg~\citeyear{Feinberg04,Feinberg05} also studied
games with awareness. Feinberg~\citeyear{Feinberg05} gives a
definition of extended Nash equilibrium in normal-form games.
Although his definition stems from much the same intuitions as ours%
\fullv{ (although some details are different---see
Section~\ref{sec:liter})}, it is expressed syntactically.  Each
player is characterized by a complete description of what moves and
players he is aware of, what moves and players he is aware that each
other player is aware of, and so on through all levels of iterated
awareness.%
\commentout{He does not give a definition of Nash equilibrium for
extensive-form games (he leaves dealing with extensive-form games as
an open problem) and he does not deal with awareness of unawareness.
Feinberg~\citeyear{Feinberg04} deals with extensive-form games, but
defines a solution concept (a generalization of \emph{sequential
equilibrium} \cite{KW82}) only indirectly, via a syntactic epistemic
characterization.}
Feinberg~\citeyear{Feinberg04} deals with extensive-form games and
defines solution concepts only indirectly, via a syntactic epistemic
characterization.
His approach lacks a more direct semantic
framework, which our model provides.  He also
does not deal with awareness of unawareness.

\fullv{The rest of this paper is organized as follows. In
Section~\ref{sec:unaware}, we describe how we represent different
awareness levels in a game. In Section~\ref{sec:equil}, we use our
representation to define a generalized notion of Nash equilibrium,
and we prove its existence in games with awareness. In
Section~\ref{sec:unawareness}, we describe how we can extend our
approach to deal with awareness of unawareness.
In Section~\ref{sec:CK}, we describe how to extend our framework to
deal with games where there is lack of common knowledge, even if
awareness is not an issue.
%
We compare our work to others in the literature, particularly
Feinberg's, in Section~\ref{sec:liter}, and
conclude in Section~\ref{sec:conc}.}

\section{Modeling awareness}
\label{sec:unaware}

The first step in dealing with awareness is modeling it.  To this
end, we consider \emph{augmented games}.  We start with a standard
game, described by a game tree $\Gamma$  (as in
Figure~\ref{fig:game1}).  An augmented game $\Gamma^+$ \emph{based
on} $\Gamma$ essentially augments $\Gamma$ by describing each
agent's \emph{awareness level} at each node, where player $i$'s
awareness level at a node $h$ is essentially the set of runs (complete
histories) in $\Gamma$ that $i$ is aware of at node $h$. A player's
awareness level may change over time,
as the player becomes aware of more moves.%

\commentout{
In standard games, we allow moves of nature to model certain types of
uncertainty.  In augmented games, we need to capture uncertainty
regarding the awareness level of other players.  For example, agent
$i$ may be aware that agent $j$ can make a certain move $m$, but may not
know whether $j$ is aware of $m$.  Or agent $i$ may know that originally
$j$ is not aware of $m$, but that, after making a certain move, there is
some probability that $j$ will become aware of $m$.  To capture this
uncertainty, we introduce a new player called the \emph{environment}.
Like nature, the environment makes randomized moves, but these moves
change only the awareness level of players.%
\footnote{We could capture all this uncertainty without introducing
the environment, by allowing nature to change the awareness level of
players.  However, it is conceptually cleaner to separate the role
of nature and the environment.}

The next step is to define the notion of {\em restricted game based
on $\Gamma$ for an awareness level $a$ of player $i$} that
represents the underlying standard game $i$ is aware of given her
awareness level. A restricted game is a standard extensive game. To
have a different underlying game for each awareness level of each
agent turns out to be necessary if for example we want to model a
situation where besides having different moves available at
histories $h$ and $h'$, player $i$ cannot distinguish those
histories because she is aware of the same set of moves at them.
Also since agents might not be aware of all moves of nature, they
may have different beliefs regarding natures' moves, we model that
in the restricted games.

We define augmented and restricted games formally in Subsections
\ref{sec:augm} \ref{sec:rest}, respectively, and then show how we
can model a game with unawareness by using a collection of augmented
games, one for the modeler, that is based on $\Gamma$, and one for
each awareness level of each agent, that are based on their
respective restricted games.

\subsection{Augmented Games}
\label{sec:augm}
}

Our formal definition of augmented game is based on
the definition of extensive game given by Osborne and
Rubinstein \citeyear{OR94}.
We start by briefly reviewing Osborne and Rubinstein's definition.

A \emph{(finite) extensive game} is a tuple $(N,M,
\H,P,f_c,\{\I_i:i\in
N\},\{u_i:i\in N\})$,
 where
\begin{itemize}
\item $N$ is a finite set consisting of the players of
the game.
\item $M$ is a finite set whose elements are the
moves \fullv{(or actions)} available to players (and nature) during the game.%
\footnote{Osborne and Rubinstein did not make $M$ explicit in their
definition of an extensive game; we find it convenient to
make it explicit here.}%
\item $\H$ is a finite set of finite sequences of moves (elements of
$M$) that is closed under prefixes, so that if $h \in \H$ and $h'$
is a prefix of $\H$, then $h' \in \H$.
%
Intuitively, each member of $\H$ is a \emph{history}.  We can identify
the nodes in a game tree with the histories in $\H$.  Each node $n$ is
characterized by the sequence of moves needed to reach $n$.
A \emph{run} in $\H$ is a terminal history, one that is not
a strict prefix of any other  history in $\H$.  Let $Z$ denote the set
of runs of $\H$.
Let $M_h = \{m \in M : h \cdot \<m\> \in \H\}$ (where we use $\cdot$
to denote concatenation of sequences); $M_h$ is the set of moves
that can be made after history~$h$.

\item $P:(\H-Z)\rightarrow N\cup\{c\}$ is a function that assigns to
each nonterminal history $h$
a member of $N\cup\{c\}$.
(We can think of $c$ as representing nature.)
If $P(h)=i$, then player $i$ moves after history $h$; if
$P(h)=c$, then nature moves after $h$.
Let $\H_i=\{h:P(h)=i\}$ be the
set of all histories after which player $i$ moves.

\item $f_c$ is a function that associates with every history for
which $P(h)=c$ a probability measure $f_c(\cdot \mid h)$ on $M_h$.
Intuitively, $f_c(\cdot\mid h)$ describes the probability
of nature's moves once history $h$ is
reached.

\item $\I_i$ is a partition of
$\H_i$
with the property that if $h$ and $h'$ are in the same cell of the
partition then $M_h=M_{h'}$, i.e., the same set of moves is
available at every history in a cell of the partition.
Intuitively, if $h$ and $h'$ are in the same cell of $\I_i$, then $h$ and
$h'$ are indistinguishable from $i$'s point of view; $i$
considers history $h'$ possible if the actual history is $h$, and vice
versa.
A cell
$I\in\I_i$ is called an ($i$-)\emph{information set}.
\item $u_i:Z\rightarrow \mathrm{R}$ is a payoff function for player $i$,
assigning a real number ($i$'s payoff) to each run of the game.
\end{itemize}

In the game of Figure~\ref{fig:game1},
\begin{itemize}
\item $N=\{A,B\}$,
$\H=\{\<\,\>, \<$down$_A\>$, $\<$across$_A$,\,down$_B\>$,
$\<$across$_A$,\,across$_B\>\}$,
\item $P(\<\,\>)=A$, $P(\<$across$_A\>)=B$,
\item $I_A=\{\<\,\>\}$, $I_B=\{\<\mbox{across}_A\>\}$,
\item $u_A(\<$down$_A\>)=u_B(\<$down$_A\>)=1$,
\item $u_A(\<$across$_A$,\,across$_B\>)=0$, and
\item $u_B(\<$across$_A$,\,across$_B\>)=2$.
\end{itemize}

In this paper, as in most work in game theory, we further assume that players
have \emph{perfect recall}:  they remember all the
actions that they have performed and the information sets they passed
through.  Formally, we require that
\begin{itemize}
\item if $h$ and $h'$ are in the same
$i$-information set and $h_1$ is a prefix of $h$ such that $P(h_1) =
i$, then there is a prefix $h_1'$ of $h'$ such that $h_1$ and $h_1'$ are
in the same information set; moreover, if $h_1 \cdot \<m\>$ is a prefix
of $h$ (so that $m$ was the action performed when $h_1$ was reached in
$h$) then $h_1' \cdot \<m\>$ is a prefix of $h'$.
\end{itemize}


An \emph{augmented game} is defined much like an extensive game; the
only essential difference is that at each nonterminal history we not
only determine the player moving but also her awareness level. Since
the awareness level is a set of runs in a game $\Gamma$, we say that
$\Gamma^+=(N^+, M^+, \H^+, P^+,f_c^+,\{\I_i^+:i\in
N^+\},\{u_i^+:i\in N^+\},\{\A_i^+ :i\in N^+\})$ is an
\emph{augmented game based on the (standard) extensive game}
$\Gamma=(N,M, \H,P,f_c,\{\I_i:i\in N\},\{u_i:i\in N\})$ if the
following conditions are satisfied:
\begin{itemize}
\item[A1.] $(N^+, M^+,\H^+, P^+,f_c^+,\{\I_i^+:i\in N^+\},\{u_i^+:i\in
N^+\})$ is a \fullv{(standard) }finite extensive game \fullv{where
players have}\shortv{with} perfect recall.
\item[A2.] $\A_i^+:\H_i^+\rightarrow 2^\H$ describes $i$'s awareness
level at each nonterminal history after which player $i$ moves. For
each $h \in \H_i^+$, $\A_i^+(h)$ consists of a set of histories in
$\H$ and all their prefixes. Intuitively, $\A_i^+(h)$ describes the
set of histories of $\Gamma$ that $i$ is aware of at history $h \in
\H_i^+$.
(Having $\A_i^+(h)$ consist of histories rather than just runs makes it
easier to deal with awareness of unawareness.)
\item[A3.] $N^+ \subseteq  N$.
\item[A4.] If $P^+(h) \in N^+$, then $P^+(h)=P(\overline{h})$, where
$\overline{h}$ is the subsequence of $h$ consisting of all the moves
in $h$ that are also in $M$, and $M^+_h\subseteq M_{\overline{h}}$.
Intuitively, all the moves available to $i$ at $h$ must also be
available to $i$ in the underlying game
$\Gamma$.

\item[A5.] If $P^+(h) = c$, then either
$P(\overline{h}) = c$ and $M^+_h \subseteq
M_{\overline{h}}$, or $M^+_h \inter M = \emptyset$. The moves in
$M^+_h$ in the case where $M^+_h \inter M= \emptyset$ intuitively
capture uncertainty regarding a player's awareness level.
%

\commentout{
\item[A6.] If $h$ and $h'$ are in the same information set in
$\I_i^+$, then $\A_i^+(h) = \A_i^+(h')$. Note that we do not require
that $\overline{h}$ and $\overline{h}'$ be in the same information
set of $\Gamma$ here. Thus, we are allowing for the possibility that
the only reason that $\overline{h}$ and $\overline{h}'$ are in
different information sets of $\Gamma$ is because there are
different moves available to $i$ at these histories.  In $\Gamma^+$,
the same moves are available to $i$ at $h$ and $h'$, so they cannot
be distinguished.
}

\item[A6.] If $h$ and $h'$ are in the same information set in
$\I_i^+$, then $\A_i^+(h) = \A_i^+(h')$.
Intuitively, $i$'s awareness level depends only on the information that
$i$ has.

\commentout{
\item[A7.] If $\A_i^+(h) = a$, then there exists some $h'$ in the same
information set as $h$ such that $\overline{h}' \in a$. Intuitively,
$i$ must consider a history in the underlying game possible that is
compatible with where he is in the augmented game.  Note that we do
not require $\overline{h}$ to be in $a$.  A move could have been in
$h$ that $i$ was not aware of, but in that case, $i$ would have to
consider it possible that another move that he was aware was made.}

\item[A7.] If $h$ is a prefix of $h'$ and $P^+(h) = P^+(h')$, then
$\A_i^+(h) \subseteq \A_i^+(h')$. This is a perfect recall
requirement; players do not forget histories that they were aware
of.

\item[A8.]
If $h$ and $h'$ are in the same information set in $\Gamma^+$,  then
 $\overline{h}$ and $\overline{h}'$ are in the same information set
in~$\Gamma$.

\item[A9.] If $h$ and $h'$ are histories in both $\Gamma^+$ and $\Gamma$%
, and
$\overline{h}$ and $\overline{h}'$ are in the same information set
in $\Gamma$, then $h$ and $h'$ are in the same information set in
$\Gamma^+$.

\item[A10.] For all $i\in N^+$ and $h\in \H_i^+$, if $h',h''\in
\A_i(h)$, $h'$ and $h''$ are in the same information set in
$\Gamma$, then $h'\cdot\<m\>\in \A_i(h)$ iff $h''\cdot\<m\>\in
\A_i(h)$.

\item[A11.]
$\{\overline{z}: z \in Z^+\} \subseteq Z$; moreover, for all $i\in
N^+$, $h\in \H^+_i$, if $z$ is a terminal history in $\A_i^+(h)$
(i.e., if $z\in A_i^+(h)$ and $z$ is not a strict prefix of another
element of $A_i^+(h)$), then $z\subseteq Z$. Thus,
the runs in $Z^+$ correspond to runs in $Z$, and players understand this
fact.

\item[A12.] For all $i \in N^+$ and runs $z$ in $Z^+$,
if $\overline{z} \in Z$, then
$u_i^+(z) =
u_i(\overline{z})$. Thus, a player's utility just depends on the
moves made in the underlying game.
(By A11, we have $\overline{z} \in Z$.  We have included the clause
``if $\overline{z} \in Z$'' so that A12 is applicable when we
consider awareness of unawareness, where we drop A11.)

\commentout{
\item[A11.]
If $h$ and $h'$ are histories in $\H^+$
such that  $P^+(h) = P^+(h') \in N^+$ and there exists histories
$h_1$, $h_1'$, and $h''$  such that (a) $h = h_1 \cdot h''$, (b) $h'
= h_1' \cdot h''$, and either (c1) $h_1 = h_1' = \<\,\>$ or  (c2)
$P^+(h_1) = P^+(h_1') = P^+(h)$ and $h_1$ and $h_1'$ are in the same
information set, then $h$ and $h'$ are in the same information set
in $\Gamma^m$. (This condition is discussed below.) }
\end{itemize}

\commentout{ The conditions A1--A8 are intended to capture our
intuitions regarding awareness.  While we found it nontrivial to get
what we considered the appropriate conditions, the exact conditions
are not critical for our results.  For example, we could drop or
weaken any of conditions A4--A8 or strengthen A6 to require that
$\overline{h}$ and $\overline{h}'$ be in the same information set
without affecting our constructions or results.}
%
Conditions A1--A12 are intended to capture our intuitions regarding
information sets, awareness, and common knowledge.
To allow us to focus on issues
directly related to awareness,
we have implicitly assumed that there is common knowledge of
(1) who moves
at histories in the underlying game (this is captured by the fact that
$P^+(h) = P(\overline{h})$ unless $P^+(h) = c$ and
$M^+_h \inter
M = \emptyset$%
\fullv{
---either the same player or nature moves at both $h$ and
$\overline{h}$ unless nature makes an ``awareness'' move at
$h$}),  (2) what the payoffs are in the underlying game (since
$u^+(\overline{z}) = u(z)$), and (3) what the information sets are in
the underlying game (see A8--A10).
\fullv{ Our approach is flexible enough to allow us to drop these
assumptions; see Section~\ref{sec:CK}.}
%
\shortv{ Our approach is flexible enough to allow us to drop these
assumptions; see the full paper for details.}

%

\fullv{To understand A8--A10, we must first discuss our view of
information sets. As pointed out by Halpern~\citeyear{Hal15},
special attention must be given to the interpretation of information
sets in game trees. This issue requires even more care in games with
awareness. The standard intuition for information sets is that a
player considers all the histories in his information set possible.
But this intuition does not apply in augmented games. In an
augmented game, there may be some histories in an $i$-information
set that
$i$ is not aware of; player $i$ cannot consider these histories
possible.  For example, consider finitely repeated prisoners dilemma
where Alice and Bob each move twice before their moves are revealed.
Even if Bob is not aware of defection, his information set after
Alice's first move in the modeler's game will still contain the
history where Alice defects.

We interpret an $i$-information set to be the set of all histories
where player $i$ has the same \emph{local state}. Intuitively, this
local states encodes all the information that $i$ has about the
moves he can make, what moves have been made, the other players in the
game, his strategy, and so on. We assume that player $i$'s local
state is characterized by the sequence of signals that that $i$ has
received in the course of the game. Therefore, $h$ and $h'$ are in
the same $i$-information set in $\Gamma$ iff $i$ received the same
sequence of signals in both histories.

In standard extensive games, the sequence of signals a player
receives after every history $h$ is assumed to be common knowledge.
(This assumption is implicit in the assumption that the game, is
common knowledge, and hence so are the information sets.)
As we said, we continue to assume this in games with awareness
(although we show how the assumption can be dropped in Section~\ref{sec:CK}).
That is why we require in
A8 that if $h$ and $h'$ are in the same $i$-information in an
augmented game, then $\overline{h}$ and $\overline{h}'$ must be in
the same $i$-information set in the underlying game. The converse of
A8 does not necessarily hold.  It could well be the case that
$\overline{h}$ and $\overline{h}'$ are in the same $i$-information
set, but since $i$ receives different signals from nature, $h$ and
$h'$ are not in the same information set. On the other hand, if all
the moves in $h$ and $h'$ are already in the underlying game, then
if $h$ and $h'$ are in the same information set of $\Gamma$, they
should be in the same information set of $\Gamma^+$. This is the
content of A9.
Since, the signals received by a player determine the moves he has
available, if player $i$ is aware of two histories in the same
information set in $\Gamma$,
he must be aware of the same set of moves available at both of these
histories. A10 captures that intuition.}

\shortv{To fully understand A8--A10, we must first discuss our view
of information sets. As pointed out by Halpern~\citeyear{Hal15},
special attention must be given to the interpretation of information
sets in game trees. This issue requires even more care in games with
awareness. The standard intuition for information sets is that a
player considers all the histories in his information set possible.
But this intuition does not apply in augmented games. In an
augmented game, there may be some histories in an $i$-information
set that
$i$ is not aware of; player $i$ cannot consider these histories
possible. We interpret an $i$-information set to be the set of all
histories where player $i$ receives the same sequence of signals.
A8-A10 reflect this interpretation under our assumption that (modulo
awareness) information sets are common knowledge.}

\commentout{
While we found it nontrivial to get what we considered the
appropriate conditions, the exact conditions are not critical for
our results.  For example, we could drop or weaken any of conditions
A4--A12 without affecting our constructions
or results.
Notice that we do not place any requirements on the utilities in an
augmented game.  This is in part to model lack of common knowledge
of the utilities, and also in part to allow the utilities to depend
on when agent's become aware of moves.  Again, our results would
continue to hold if we required, for example, that $u^+(z) =
u(\overline{z})$.
}

For the remainder of the paper, we use the following notation: for a
(standard or augmented) game $\Gamma^s$, we denote the components of
$\Gamma^s$ with the same superscript $s$, so that we have $M^s$,
$\H^s$, and so on. Thus, from here on we do not explicitly describe
the components of a game.

\begin{figure}[htb]
\centering \epsfxsize=14cm \epsffile{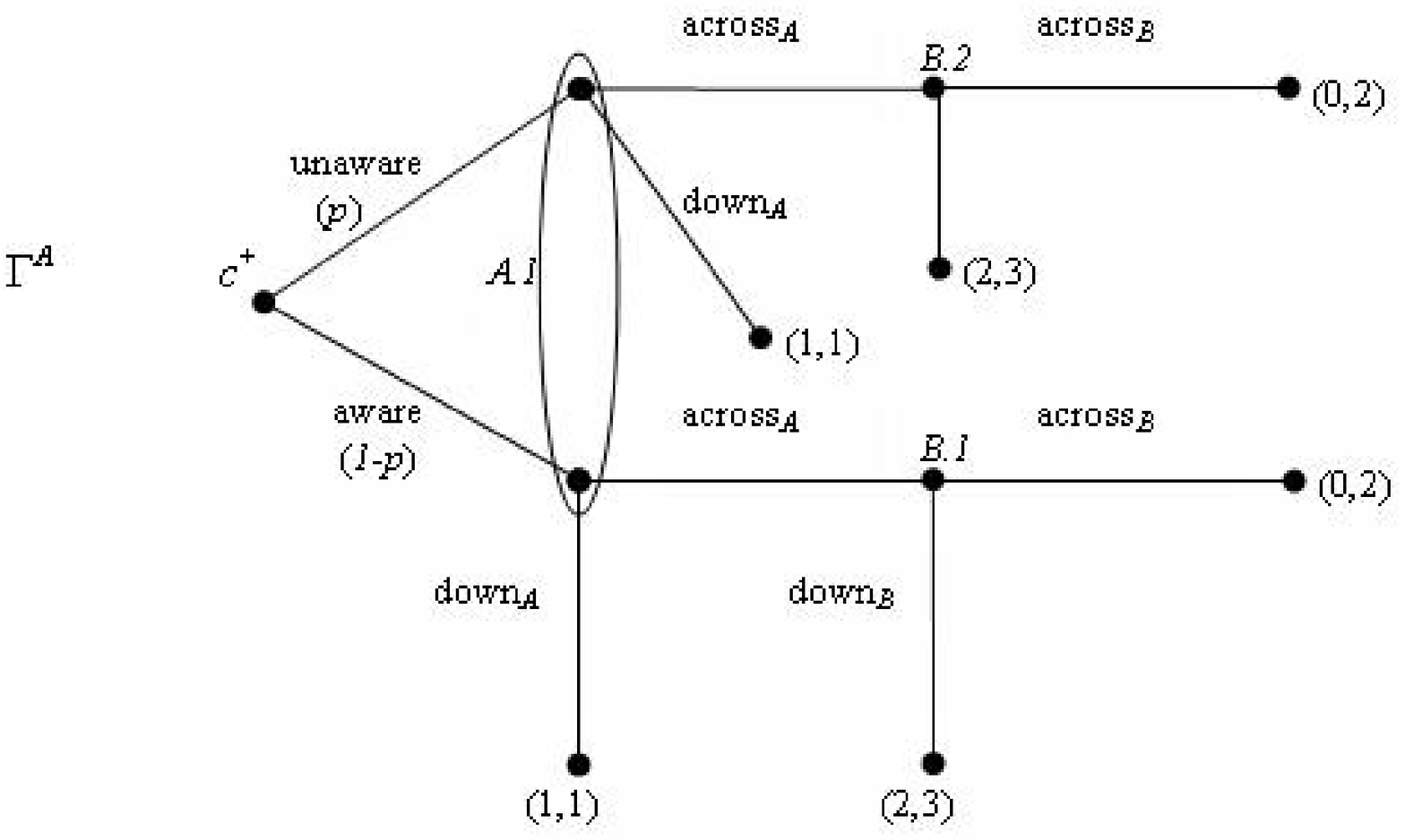} \caption{The
augmented game $\Gamma^A$.} \label{fig:A1game}
\end{figure}

An augmented game describes either the modeler's view of the game or the
subjective view of the game of one of the players, and includes both
moves of the underlying game and
moves of nature that change awareness.
For example, consider again the game shown in Figure~\ref{fig:game1} and
suppose that
\begin{itemize}
\item players $A$ and $B$ are aware of all histories of the game;
\item player $A$ is uncertain as to  whether player $B$ is aware of run
$\<$across$_A$,\,down$_B\>$ and believes that he is unaware of it
with probability $p$; and
\item the type of player $B$ that is
aware of the run $\<$across$_A$, down$_B\>$ is aware that player $A$
is aware of all histories, and he knows $A$ is uncertain about his
awareness level and knows the probability $p$.
\end{itemize}
Because $A$ and $B$ are actually aware of all histories of the underlying
game, from the point of view of the modeler, the augmented game is
essentially identical to the game described in
Figure~\ref{fig:game1}, with the awareness level of both players $A$
and $B$ consisting of all histories of the underlying game. However, when
$A$ moves at the node labeled $A$ in the modeler's game, she
believes that the actual augmented game  is $\Gamma^A$, as described
in Figure~\ref{fig:A1game}.  In $\Gamma^A$, nature's initial move
captures $A$'s uncertainty about $B$'s awareness level. At the
information set labeled $A.1$, $A$ is aware of all the runs of the
underlying game. Moreover, at this information set, $A$ believes
that the true game is $\Gamma^A$.

\begin{figure}[htb]
\centering \epsfxsize=10cm \epsffile{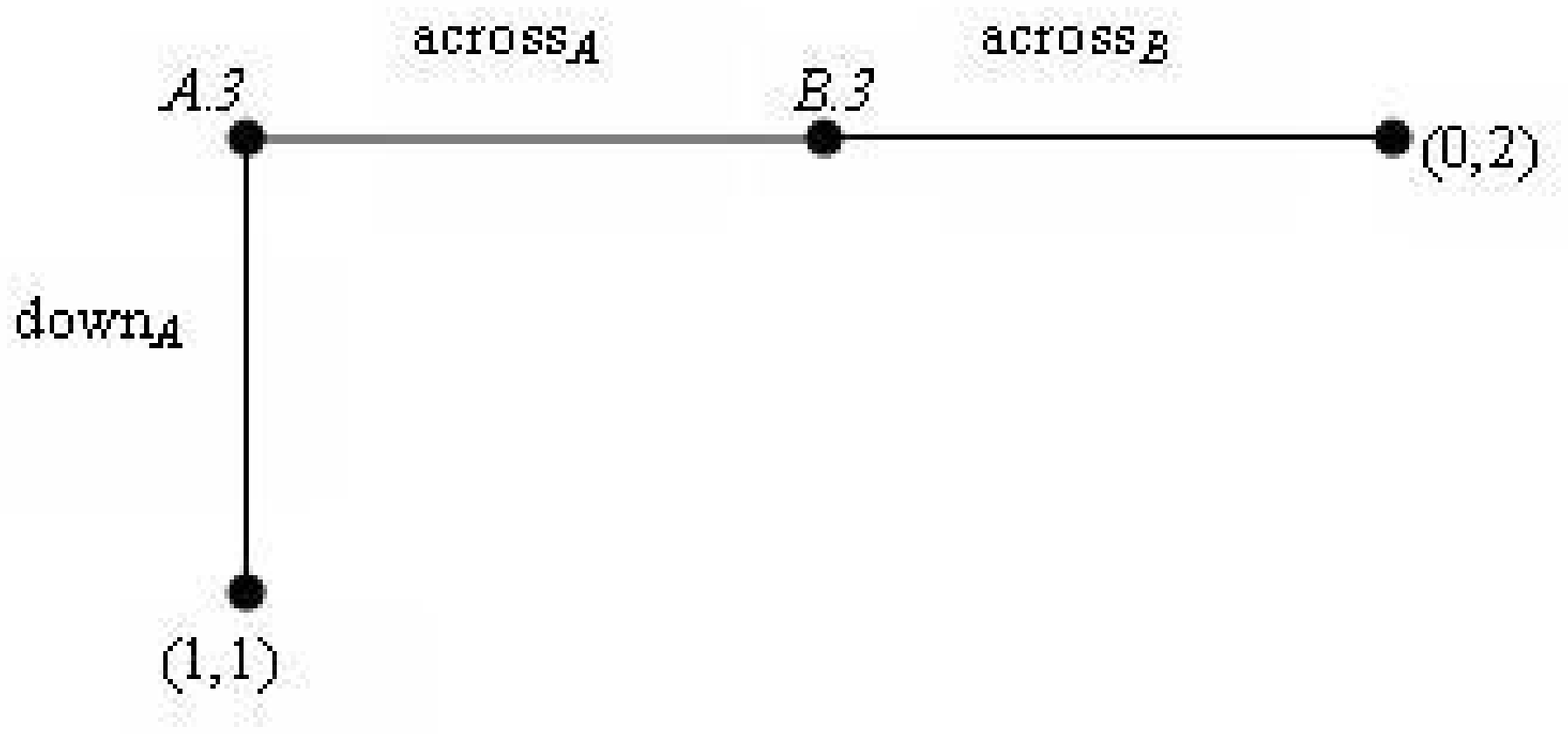} \caption{The
augmented game $\Gamma^B$.} \label{fig:A2game}
\end{figure}

At the node labeled $B.1$, $B$ is aware of all the runs of the
underlying game and believes that the true game is the modeler's
game; but at the node labeled $B.2$, $B$ is not aware that he can
play down$_B$, and so believes that the true game is the augmented
game $\Gamma^B$ described in Figure~\ref{fig:A2game}. At the nodes
labeled $A.3$ and $B.3$ in the game $\Gamma^B$, neither $A$ nor $B$
is aware of the move down$_B$.  Moreover, both players think the
true game is $\Gamma^B$.

As this example should make clear, to model a game with possibly
unaware players, we need to consider not just one augmented game,
but a collection of them.  Moreover, we need to describe, at each
history in an augmented game, which augmented game the player
playing at that history believes is the actual augmented game being
played.

To capture these intuitions, we define a \emph{game with awareness
based on $\Gamma = (N, M,\H,P,f_c, \{\I_i:i\in N\},\{u_i: i\in
N\})$} to be a tuple $\Gamma^* = (\G, \Gamma^m, \F)$, where
\begin{itemize}
\item $\G$ is a countable set of augmented games based on $\Gamma$, of
which one is
$\Gamma^m$;
\item $\F$ maps an augmented game $\Gamma^+ \in\G$ and a history $h$ in
$\Gamma^+$ such that $P^+(h)=i$ to a pair $(\Gamma^h,I)$, where
$\Gamma^h\in \G$ and $I$ is an $i$-information set in game
$\Gamma^h$.
\end{itemize}
Intuitively, $\Gamma^m$ is the game from the point of view of an
omniscient modeler. If player $i$ moves at $h$ in game $\Gamma^+ \in
\G$ and $\F(\Gamma^+,h) = (\Gamma^h,I)$, then $\Gamma^h$ is the game
that $i$ believes to be the true game when the history is $h$, and $I$
consists of the set of histories  in $\Gamma^h$ he currently
considers possible. For example, in the examples described in
Figures~\ref{fig:A1game} and~\ref{fig:A2game}, taking $\Gamma^m$ to
be
the augmented game in Figure~\ref{fig:game1}, we have
$\F(\Gamma^m,\<\, \>) = (\Gamma^A,I)$, where $I$ is the information
set labeled $A.1$ in Figure~\ref{fig:A1game}, and
$\F(\Gamma^A,\<$unaware,across$_A\>) =
(\Gamma^B,\{\<$across$_A\>\})$.

It may seem that by making $\F$ a function we cannot capture a
player's uncertainty about the game being played.  However, we can
capture such uncertainty by folding it into nature's move.  For
example,
we capture $A$'s uncertainty about whether $B$ is aware of being able to
move
down$_B$ in the augmented game $\Gamma^A$ illustrated in
Figure~\ref{fig:A1game} by having nature decide this at the first
step. It should be clear that this gives a general approach to
capturing such uncertainty.

\commentout{ The $\F$ function allows us to capture the fact that
the underlying game may not be common knowledge.  For example, we
allow the utilities of a run $u^+(z)$ of an augmented game
$\Gamma^+$ to be different from $\u(\overline{z})$ since we want to
allow uncertainty about the utility function.
Similarly, as we shall see, we allow players to be uncertain about
the probabilities of nature's moves in the underlying game and about
what the information sets in the underlying game are.

Nevertheless, }

The augmented game $\Gamma^m$ and the mapping $\F$ must satisfy a
number of consistency conditions. The first set of conditions
applies to $\Gamma^m$.  Since the modeler is presumed to be
omniscient, the conditions say that the modeler is aware of all the
players and moves of the underlying game.
\begin{itemize}
\item[M1.] $N^m=N$.
\item[M2.] $M \subseteq M^m$ and $\{\overline{z}:z\in Z^m\}=Z$.
\item[M3.]
If $P^m(h) \in N$, then $M^m_h = M_{\overline{h}}$.
If $P^m(h) = c$, then
either
$M^m_h \inter M = \emptyset$ or $M^m_h = M_{\overline{h}}$ and
$\mbox{$f^m_c( \cdot \mid h)$} =  f_c(\cdot \mid \overline{h})$.

\commentout{
\item[M4.] If
$h$ and $h'$ are nonterminal histories in $\H^m$, $\overline{h}$ and
$\overline{h}'$ are in the same information set in $\Gamma$, and
$\A_i^m(h) = \A_i^m(h')$, then $h$ and $h'$ are in the same
information set in $\I_i^m$.  Con
$h=<h_{-}\cdot h_1>$ and $h'=<h'_{-}\cdot
h'_1>$
if $\overline{h}$ and
$\overline{h}'$ are in the same information set in $\Gamma$ and
$\A_i^m(h) \ne \A_i^m(h')$, then there exists a history $h'' \in
\H^m$ such that $\overline{h}'' = \overline{h}'$ and
$h$ and $h''$ are in the same information set in $\Gamma^m$.%
}

\commentout{
\item[M4.]
If $h$ and $h'$ are histories in the same information set in
$\Gamma^m$,  then
 $\overline{h}$ and $\overline{h}'$ are in the same information set
in~$\Gamma$.
\item[M5.] If $h$ and $h'$ are histories in both $\Gamma^m$ and $\Gamma$
(i.e., $h = \overline{h}$ and $h' = \overline{h}'$), and
$\overline{h}$ and $\overline{h}'$ are in the same information
set in $\Gamma$, then $h$ and $h'$ are in the same information set
in $\Gamma^m$.

}
\end{itemize}

M1, M2 and M3 enforce the intuition that the modeler understands the
underlying game.  He knows all the players and possible moves, and
understands how nature's moves work in the underlying game $\Gamma$.
It may seem somewhat surprising that there is no analogue of the
second part of M3 (i.e., the constraint of $f_c^m$) for all
augmented games. While it makes sense to have such an analogue if
nature's moves are in some sense objective, it seems like an
unreasonable requirement that all player's should agree on these
probabilities in general.  This is especially so in the case that a
player suddenly becomes aware of some moves of nature that he was
not aware of before.  It does not seem reasonable to assume that
this awareness should come along with an understanding of the
probabilities of these moves.  Of course, we could require
such
an
analogue of M3.  Since the set of games that have such a requirement
is a subset of the games we consider, all our results apply without
change if such a requirement is imposed.
\commentout{M4 enforces the intuition that if  player $i$ cannot
distinguish between histories $\overline{h}$ and $\overline{h}'$ in
$\Gamma$, then $i$ should not be able to distinguish between $h$ and
$h'$ in $\Gamma^m$. This is not quite true, because $i$ may have
different awareness levels at $h$ and $h'$, in which case $h$ and
$h'$ cannot be in the same information set. However, there should
exist a history $h''$ with the same underlying history such that $h$
and $h''$ are in the same information set in $\Gamma^m$. However,
this should no longer be interpreted as saying that $i$ cannot
distinguish $h$ and $h''$, since $i$ may not be aware of all the
moves in $h''$.  A better interpretation might be that $i$ gets the
same signals in $h$ and $h''$.
}%

\commentout{
To understand M4 and M5, we must first discuss our view
of information sets. As pointed out by Halpern~\citeyear{Hal15},
special attention must
be given to the interpretation of information sets in game
trees. This issue requires even more care in games with awareness.
The standard intuition for information sets is that a player
considers all the histories in his information set possible.  But
this intuition does not apply in augmented games.
In an augmented game, there may be some histories in
an $i$-information set that include moves of which $i$ is not
aware; player $i$
cannot consider these histories possible.  For example, consider
finitely repeated prisoners dilemma where Alice and Bob each move twice
before their moves are revealed.  Even if Bob is not aware of defection, his
information set after Alice's first move in the modeler's game
will still contain the history where Alice defects.

We interpret an $i$-information set to be the set
of all histories where player $i$ has the same \emph{local
state}. Intuitively,
this local states encodes all the information that $i$ has about the
moves he can make, what moves have been made, the other
players in the game, his strategy, and so on.
We assume that player $i$'s local state is characterized by the sequence
of signals that that $i$ has received in the
course of the game. Therefore, $h$ and $h'$ are in the same
$i$-information set in $\Gamma$ iff $i$ received
the same sequence of signals in both histories.

In standard extensive games, the sequence of signals a player
receives after every history $h$ is assumed to be common knowledge.
(This assumption is implicit in the assumption that the game,
is common knowledge, and hence so are the information sets.)
We do not assume this in games with awareness.  In particular,
although we assume that each player $i$ knows what signals he gets
in each history, we allow a player $j$ to have false beliefs about
the signals that another player $i$ gets. That is why we do not
require in A6 that if $h$ and $h'$ are in the same information set
of player $i$ in an augmented game, then $\overline{h}$ and
$\overline{h}'$ must be in the same information set in the
underlying game.  We would want to make this requirement if the
signals that $i$ receives in a history of the underlying game are common
knowledge (or, equivalently, if the partition of the underlying game
into information sets is common knowledge).
Note that
if $\Gamma^+$ is the augmented
game from $j$'s point of view, then
$j$ will believe that for histories $h$ and $h'$ in the same
$i$-information set, $i$ gets the same signals in $\overline{h}$ and
$\overline{h}'$, and thus that $\overline{h}$ and $\overline{h}'$
are in the same information set in the underlying game, although $j$
might be mistaken.
We cannot express $j$'s beliefs about the information sets in the
underlying game in our framework, but we do not feel that this is a
significant deficiency.

Since we assume that the modeler understands the signals that each
player gets as a result of moves made in $\Gamma$, the modeler realizes
that if two histories in $\Gamma^m$ are in the same
$i$-information set, then $i$ must have gotten the same
signals in
$\overline{h}$ and $\overline{h}'$, so they must be in the same
information set of $\Gamma$.  This is captured by M4.
The converse of M4 does not necessarily hold.  It could well be the case
that $\overline{h}$ and $\overline{h}'$ are in the same information set
of player $i$, but since $i$ receives different signals from nature, $h$
and $h'$ are not in the same information set.
On the other hand, if all the moves in $h$ and $h'$ are already in the
underlying game, then if $h$ and $h'$ are in the same information set of
$\Gamma$,
they should
be in the same information set of $\Gamma^m$. This is the content of
M5.
}


Although the modeler understands the underlying game $\Gamma$,
$\Gamma^m$ is not uniquely determined by $\Gamma$.  There may be many
modeler's games based on $\Gamma$, where the players have different
awareness levels and the awareness changes in different ways.

The game $\Gamma^m$ can be thought of as a description of
``reality''; it describes the effect of moves in the underlying game
and how players' awareness levels change. The other games in $\G$
describe a player's subjective view of the situation.
The constraints on the mapping $\F$ that we now describe capture
desirable properties of awareness.

\commentout{
To explain these constraints, we need one more definition.
Let $\G_{m,i}$ be the smallest subset of $\G$ such that
if either $\Gamma^+ = \Gamma^m$ or $\Gamma^+ \in \G_{m,i}$,
$h\in H^+$, $P^+(h)=i$, and
$\F(\Gamma^+,h)=(\Gamma',\cdot)$, then $\Gamma'\in\G_{m,i}$.
Intuitively, $\G_{m,i}$ consists of all games player $i$ considers
possible, or considers possible that he considers possible, and so
on, at some history of the modeler's game.
}

Consider the following constraints, where
$\Gamma^+\in \G$, $h
\in \H^+$, $P^+(h) = i$, $\A_i^+(h) = a$, and $\F(\Gamma^+,h) =
(\Gamma^h,I)$.
\begin{itemize}

\item[C1.]$\{\overline{h}: h \in \H^h\} = a$.

\item[C2.] If $h' \in \H^h$ and $P^h(h') = j$, then $\A_j^h(h')
\subseteq a$ and $M_{\overline{h}'}\inter \{m: \overline{h}'\cdot\<
m\>\in a\}= M^h_{h'}$.

\item[C3.] If $h'$ and $h$ are in the same information set in
$\Gamma^+$ and $\overline{h}' \in a$, then there exists $h'' \in I$
such that $\overline{h}''=\overline{h}'$.

\item[C4.] If $h'\in I$, then $\A_i^h(h')=a$ and
$\F(\Gamma^h,h')=(\Gamma^h,I)$.

\item[C5.] If $h' \in \H^+$, $P^+(h') = i$, $\A_i^+(h')=a$,
then if
$h$ and $h'$ are in the same information set of $\Gamma^+$, then
$\F(\Gamma^+,h') = (\Gamma^{h},I)$, while
if $h$ is a prefix or a suffix of $h'$, then
$\F(\Gamma^+,h') = (\Gamma^{h},I')$ for some $i$-information set $I'$.


\commentout{
\item[C6.]
If $\Gamma^h \in \G_{m,i}$, then
\begin{itemize}
\item[(a)]
if $h' \in I$, then
$\overline{h}$ and $\overline{h}'$ are in the same information set in
$\Gamma$;
\item[(b)] if $h'$ and $h''$ are histories in the same $i$-information
set in $\Gamma^h$, then $\overline{h}'$ and $\overline{h}''$ are in
the same $i$-information set in $\Gamma$; and
\item[(c)] if $h'$ and $h''$ are histories in both $\Gamma^h$ and
$\Gamma$ and are in the same $i$-information set in $\Gamma$, then they
are in the same $i$-information set in $\Gamma^h$.
\end{itemize}
}
\item[C6.]
If $h' \in I$, then $\overline{h}$ and $\overline{h}'$ are in the
same information set in $\Gamma$;

\item[C7.] If $\Gamma^h = \Gamma^+$, then
$h'\in I$ iff $h$ and $h'$ are in the same $i$-information set in
$\Gamma^+$.

\item[C8.] For all histories $h'\in I$, there exists a prefix $h'_1$ of
$h'$ such
that $P^h(h'_1)=i$ and $\F(\Gamma^h,h'_1)=(\Gamma',I')$ iff there
exists a prefix $h_1$ of $h$ such that $P^+(h_1)=i$ and
$\F(\Gamma^+,h_1)=(\Gamma',I')$.
Moreover, $h'_1\cdot \<m\>$ is a
prefix of $h'$ iff $h_1\cdot \<m\>$ is a prefix of $h$.

\item[C9.] There exists
a history
$h'\in I$ such that for every prefix
$h''\cdot \<m\>$ of $h'$, if $P^h(h'')=j \in N^h$
and $\F(\Gamma^h,h'')=(\Gamma',I')$, then for all $h_1\in I'$,
$h_1\cdot \<m\>\in \H'$.

\item[C10.] If $h'$ and $h''$ are histories in both $\Gamma^+$ and
$\Gamma^h$, then $h'$ and $h''$ are in the same $i$-information set
in $\Gamma^+$ iff $h'$ and $h''$ are in the same $i$-information set
in $\Gamma^h$.

\end{itemize}

Suppose that $\F(\Gamma^+,h) = (\Gamma^h,I)$. Player $i$ moving at
history $h$ in $\Gamma^+$ thinks the actual game is $\Gamma^h$.
Moreover, $i$ thinks he is in the information set of $I$ of
$\Gamma^h$.
C1 guarantees that the set of histories of the underlying game
player $i$ is aware of is exactly the set of histories of the
underlying game that appear in $\Gamma^h$.
C2 states that no player in $\Gamma^h$ can be aware of histories not
in $a$. The second part of C2 implies that the set of moves
available to player $j$ at $h'$ is just the set of moves that
player $i$ is aware of that are available to $j$ at $\overline{h}'$
in the underlying game.
C3 guarantees that for all histories $h'$ indistinguishable from $h$
that player $i$ is aware of, there exists some history $h''\in I$
differing from $h'$ at most in some moves of nature that change
awareness levels.
%
C4 says that at all histories in $I$ player $i$ indeed thinks the
game is $\Gamma^h$ and that the information set is $I$.
C5 says that player $i$'s subjective view of the game changes only
if $i$ becomes aware of more moves and is the same at histories in
$\H^+$ that $i$ cannot distinguish.
\commentout{
C6(a) captures the assumption that at all histories $i$
considers possible, he must have gotten the same signals as he does
in the actual history.  Parts (b) and (c) of C6 are analogues of M4
and M5.
Note that no part of C6 necessarily holds in a game
$\Gamma'$ that $j$ considers possible that $i$ considers possible,
because in $\Gamma'$, $j$ may have an imperfect understanding of
$i$'s interpretation of the signals in the underlying game, and
$\Gamma'$ ultimately depends on $j$'s subjective understanding of
the game.
(Again, $j$ will believe that all these conditions hold, but we have no
way of describing $j$'s beliefs about the information sets in the
underlying game.)
Also note that if $\Gamma^h \in \G_{m,i}$, then either
$\Gamma^+ \in
\G_{m,i}$ or $\Gamma^+ = \Gamma^m$, so by C6(b) and M4, it follows
that for any histories $h_1$ and $h_1'$ in the same $i$-information
set in $\Gamma^+$, it must be the case $\overline{h}_1$ and
$\overline{h}_1'$ are in the same $i$-information set.
}
C6 captures the assumption that at all histories $i$ considers
possible, he must have gotten the same signals as he does in the
actual history.

C7 says that if while moving at history $h$ player $i$ thinks that
$\Gamma^+$ is the actual game, then he considers possible all and
only histories in the information set containing $h$. C8 is a
consequence of the perfect recall assumption. C8 says that if, at
history $h$, $i$ considers $h'$ possible, then for every prefix
$h_1'$ of $h'$ there is a corresponding prefix of $h$ where $i$
considers himself to be playing the same game, and similarly, for
every prefix of $h$ there is a prefix of $h'$ where $i$ considers
himself to be playing the same game.  Moreover, $i$ makes the same
move at these prefixes.

The intuition behind condition C9 is that player $i$ knows that
player $j$ only make moves that $j$ is aware of. Therefore, player
$i$ must consider at least one history $h'$ where he believes
that every player $j$ made a move that $j$ was aware of. It follows
from A11, C1, C2, and C9 that there is a run going through $I$ where
every player $j$
makes a move that player $i$ believes that $j$ is aware of.

Since we assume that players have (modulo awareness) common knowledge
about information sets, if $\Gamma^+$ is the game from the point of view
of player $j$ (or the modeler) and there are histories $h'$ and
$h''$ in both $\Gamma^+$ and $\Gamma^h$, then player $j$ (or the
modeler) knows that player $i$ gets the same signals in both $h'$
and $h''$ iff he knows that player $i$ knows that he gets the same
signals in those histories. C10 captures that intuition.

Just as $\Gamma^m$ is not uniquely determined by $\Gamma$,
$\F(\Gamma^+,h)$ depends on more than just the awareness level of
the player who moves at $h$. That is, even if $\A_i(h) = \A_i(h')$,
we may have
$\F(\Gamma^+,h) = (\Gamma^h,I)$ and $\F(\Gamma^+,h') =
(\Gamma^{h'},I')$ with $\Gamma^{h} \ne \Gamma^{h'}$.
\commentout{
The game that $i$ believes he is playing may depend not
only on the histories that $i$ is aware of, but also on how $i$
became aware of the histories.
\commentout{
when he became aware of these histories. For example, consider the
case of two beginners $i$ and $j$ playing chess. Suppose that
initially both players are not aware of the ``en passant'' move but
are aware of all other moves. They can learn about ``en passant''
either by reading the chess rule book they both have access to or by
observing the other player play it.
player $i$ becomes aware of the ``en passant'' move as a result of
reading about it. Player $i$ is now uncertain as to whether $j$ is
aware of the ``en passant'' move. It seems reasonable that the
probability $p$ that $i$ ascribes to $j$ being aware of the ``en
passant'' move should depend on when in the game $i$ learns about
it. Intuitively, the later $i$ becomes aware of it, the larger $p$
should be.
}
For example, the order that an agent learns about moves could have a
significant on the game that she thinks he is playing.}
We do not require that the awareness level determines the game a
player considers possible. This extra flexibility allows us to model
a situation where,
for example,
players 2 and 3, who have the same awareness level and
agree on the awareness level of player 1, have
different beliefs about the game player 1 considers
possible.%
\fullv{\footnote{If the beliefs of players 2 and 3 regarding 1 are
compatible with a common prior, then we can view players 2 and 3 as
considering different information sets in the same game possible.
However, if their beliefs are not compatible with a common prior,
for example, if player 2 believes that player 1 believes that, in
history $h$, $\Gamma_1$ is the actual game with probability 1, and
player 3 believes that, in history $h$, player 1 believes that
$\Gamma_2$ is the actual game with probability 1, where $\Gamma_1
\ne \Gamma_2$, then we cannot view players 2 and 3 as considering
the same game possible.} }

A standard extensive game $\Gamma$ can be
identified with the game $(\{\Gamma^m\},\Gamma^m,\F)$, where
(abusing notation slightly) $\Gamma^m = (\Gamma,\{\A_i: i \in N\})$
and, for all histories $h$ in an $i$-information set $I$ in $\Gamma$,
$\A_i(h) = \H$ and $\F(\Gamma^m,h) = (\Gamma^m,I)$.
Thus, all players are aware of all the runs in $\Gamma$, and agree
with each other and the modeler that the game is $\Gamma$. We call
this the \emph{canonical representation of $\Gamma$} as a game with
awareness.

One technical issue: We have assumed that the set $\G$ of games in a
game $\Gamma^*$ with awareness is countable.  For our purposes, this
is without loss of generality.  We are ultimately interested in what
happens in the game $\Gamma^m$, since this is the game actually
being played. However, to analyze that, we need to consider what
happens in other games in $\G$.  For example, if $h$ is a history in
$\Gamma^m$ where $i$ moves, we need to understand what happens in
the game $\Gamma^h$ such that $\F(\Gamma^m,h) = (\Gamma^h,\cdot)$,
since $\Gamma^h$ is the game that $i$ thinks is being played at
history $h$ in $\Gamma^m$.  It is not hard to see that the set of
games we need to consider is the least set $\G'$ such that $\Gamma^m
\in \G'$ and, for every $\Gamma' \in \G$ and history $h$ in
$\Gamma'$ such that $\F(\Gamma',h)=(\Gamma'',\cdot)$, $\Gamma'' \in
\G'$. $\G'$ is guaranteed to be countable, even if $\G$ is not.

\shortv{Finally, we remark that our notion of a game with awareness
as consisting of the modeler's game together with description of
which game each agent thinks is the actual game at each history has
much in common with the intuition behind Gal and Pfeffer's
\citeyear{GP05} notion of a \emph{Network of Influence Diagrams}; we
discuss the connections in detail in the full paper.}

\section{Local strategies and generalized Nash equilibrium} \label{sec:equil}

\subsection{Local Strategies}
In this section, we generalize the notion of Nash equilibrium to
games with awareness. To do that, we must first define what a
strategy is in a  game with awareness.
Recall that in a standard game, a strategy for player $i$ is a function
from $i$-information sets to a move or to a distribution over moves,
depending on whether we are considering \emph{pure}
(i.e., deterministic) strategies or \emph{behavioral} (i.e., randomized)
strategies.  The intuition is that
player $i$'s
actions depend on what $i$ knows; the strategy can be viewed as a
\emph{universal plan}, describing what $i$ will do in every possible
situation that can arise.  This makes sense only because $i$ is presumed
to know the game tree, and thus to know in advance all the situations that
can arise.

In games with awareness, this intuition no longer makes sense.  For
example, player $i$ cannot plan in advance for what will happen if
he becomes aware of something he is initially unaware of.  We must
allow $i$'s strategy to change if he becomes aware of more moves.
Let $\G_i=\{\Gamma'\in\G:\, \mbox{for some } \Gamma^+\in\G \mbox{ and
}h\mbox{ in }\Gamma^+, \, P^+(h)=i\mbox{ and
}\F(\Gamma^+,h)=(\Gamma',\cdot)\}$. Intuitively, $\G_i$ consists of
the games that $i$ views as the real game in some history.
%
Thus, rather than considering a single strategy in a game $\Gamma^*
= (\G, \Gamma^m,\F)$ with awareness, we consider
a collection $\{\sigma_{i,\Gamma'}:
\Gamma'\in \G_i \}$ of what we call \emph{local strategies}, one for
each augmented game in $\G_i$.  Intuitively, a local strategy
$\sigma_{i,\Gamma'}$ for game $\Gamma'$ is the strategy that $i$
would use if $i$ were called upon to play and $i$ thought that the
true game was $\Gamma'$.  Thus, the domain of $\sigma_{i,\Gamma'}$
consists of pairs $(\Gamma^+,h)$ such that $\Gamma^+ \in \G$, $h$ is
a history in $\Gamma^+$, $P^+(h) = i$, and $\F(\Gamma^+,h) =
(\Gamma',I)$.


Define an equivalence relation $\sim_i$ on pairs
$(\Gamma',h)$ such that $\Gamma' \in \G$ and $h$ is a history in
$\Gamma'$ where $i$ moves
\commentout{ such that $(h_1, \Gamma_1) \sim_i (h_2, \Gamma_2)$ if
(a) both $\F(\Gamma_1,h_1) = \F(\Gamma_2,h_2)$, (b) $i$ has the same
awareness level at $h_1$ and $h_2$, and (c) $\overline{h}_1$ is the
same information set of $\Gamma$ as $\overline{h}'$.  We can think
of $\sim_i$ as defining a \emph{generalized} information partition
in $\Gamma^*$. It is easy to check that a $\sim_i$ equivalence class
consists of a union of information sets in individual games in $\G$.
Moreover, if some element of a $\sim_i$ equivalence class is in the
domain of $\sigma^{\Gamma'}$, then so is the whole equivalence
class. At all pairs $(\Gamma',h')$ in a $\sim_i$ equivalence class,
player $i$ thinks he is actually playing in the same information set
$\F(\Gamma',h')$. Thus, we require that
$\sigma^{\Gamma'}((\Gamma_1,h_1) = \sigma^{\Gamma'}(\Gamma_2,h_2)$
if $\Gamma_1,h_1)$ and $(\Gamma_2,h_2)$ are both in the domain of
$\sigma^{\Gamma'}$ and $(h_1, \Gamma_1) \sim_i (h_2, \Gamma_2)$. }
such that $(\Gamma_1,h_1) \sim_i (\Gamma_2,h_2)$ if
$\F(\Gamma_1,h_1) = \F(\Gamma_2,h_2)$.  We can think of $\sim_i$ as
defining a \emph{generalized information partition} in $\Gamma^*$.
It is easy to check that a $\sim_i$ equivalence class consists of a
union of $i$-information sets in individual games in $\G$. Moreover, if
some element of a $\sim_i$ equivalence class is in the domain of
$\sigma_{i,\Gamma'}$, then so is the whole equivalence class. At all
pairs $(\Gamma',h')$ in a $\sim_i$ equivalence class, if
$\F(\Gamma',h')=(\Gamma^{h'},I)$, player $i$ thinks he is
actually playing in the information set $I$ of
$\Gamma^{h'}$. Thus, we require that
$\sigma_{i,\Gamma'}(\Gamma_1,h_1) =
\sigma_{i,\Gamma'}(\Gamma_2,h_2)$ if $(\Gamma_1,h_1)$ and
$(\Gamma_2,h_2)$ are both in the domain of $\sigma_{i,\Gamma'}$ and
$(\Gamma_1, h_1) \sim_i (\Gamma_2, h_2)$.

The following definition summarizes this discussion.
\begin{definition} Given a game with awareness $\Gamma^* = (\G,$ $\Gamma^m,\F)$,
a \emph{local strategy} $\sigma_{i,\Gamma'}$ for agent $i$
is a function mapping pairs $(\Gamma^+,h)$ such
that $h$ is a history where $i$ moves in $\Gamma^+$ and
$\F(\Gamma^+,h) = (\Gamma',I)$ to a probability distribution over
$M'_{h'}$, the moves available at a history $h' \in I$, such that
$\sigma_{i,\Gamma'}(\Gamma_1,h_1) =
\sigma_{i,\Gamma'}(\Gamma_2,h_2)$ if $(\Gamma_1,h_1) \sim_i
(\Gamma_2,h_2)$.
\end{definition}

Note that there may be no relationship between the strategies
$\sigma_{i,\Gamma'}$ for different games $\Gamma'$. Intuitively,
this is because discovering about the possibility of a different
move may cause agent $i$ to totally alter his strategy. We could
impose some consistency requirements,
but we have not found any that we believe should hold in all games.
We believe that all our results
would continue to hold
in the presence of reasonable additional requirements,
although we have not explored the space of such requirements.

\subsection{Generalized Nash Equilibrium}
\label{sec:gnash}

We want to define a notion of generalized Nash equilibrium so as to
capture the intuition that for every player $i$, if $i$ believes he
is playing game $\Gamma'$, then his local strategy
$\sigma_{i,\Gamma'}$ is a best response to the local strategies of
other players in~$\Gamma'$.

Define a {\em generalized strategy profile} of $\Gamma^* =
(\G,\Gamma^m,\F)$ to be a set of local strategies $\vec{\sigma} =
\{\sigma_{i,\Gamma'}:i\in N,\Gamma'\in\G_i\}$. Let
$EU_{i,\Gamma'}(\vec{\sigma})$ be the expected payoff for $i$ in the
game $\Gamma'$ given that strategy profile $\vec{\sigma}$ is used.
Note that the only strategies in $\vec{\sigma}$ that are needed to
compute $EU_{i,\Gamma'}(\vec{\sigma})$ are the strategies actually
used in $\Gamma'$; indeed, all that is needed is the restriction of
these strategies to information sets that arise in $\Gamma'$.

A \emph{generalized Nash
equilibrium} of $\Gamma^* = (\G,\Gamma^m,\F)$ is a generalized
strategy profile $\vec{\sigma}$ such that for all $\Gamma'
\in \G_i$, the local strategy $\sigma_{i,\Gamma'}$ is a best
response to $\vec{\sigma}_{-(i,\Gamma')}$, where
$\vec{\sigma}_{-(i,\Gamma')}$ is the set of all local strategies in
$\vec{\sigma}$ except $\sigma_{i,\Gamma'}$.

\begin{definition}
A generalized strategy profile $\vec{\sigma}^*$ is a
\emph{generalized Nash equilibrium} of a game $\Gamma^* =
(\G,\Gamma^m,\F)$ with awareness if, for every player $i$, game
$\Gamma' \in \G_i$, and local strategy $\sigma$ for $i$ in
$\Gamma'$,
$$EU_{i,\Gamma'}(\vec{\sigma}^*)\geq
EU_{i,\Gamma'}((\vec{\sigma}^*_{-(i,\Gamma')},\sigma)).$$
\end{definition}

The standard definition of Nash equilibrium would say that
$\vec{\sigma}$ is a Nash equilibrium if $\sigma_i$ is a best
response to $\vec{\sigma}_{-i}$.  This definition implicitly assumes
that player $i$ can choose a whole strategy.  This is inappropriate
in our setting. An agent cannot anticipate that he will become aware
of more moves. Essentially, if $\Gamma_1 \ne \Gamma_2$, we are
treating player $i$ who considers the true game to be $\Gamma_1$ to
be a different agent from the version of player $i$ who considers
$\Gamma_2$ to be the true game. To understand why this is
appropriate, suppose that player $i$ considers $\Gamma_1$ to be the
true game, and then learns about more moves, and so considers
$\Gamma_2$ to be the true game.  At that point, it is too late for
player $i$ to change the strategy he was playing when he thought the
game was $\Gamma_1$.  He should just try to play optimally for what
he now considers the true game.
Moreover, while player $i$ thinks that the game $\Gamma_1$ is the
true game, he never considers it possible that he will ever be
playing a different game, so that he cannot ``prepare himself''
for a change in his subjective view of the game.%
\footnote{ In games with awareness of unawareness, an agent
may consider it possible that he will become aware of more
information.  But this too is incorporated in his view of the game,
so he can still do no better than playing optimally in his current
view of the game.}
These considerations suggest that our notion of
Nash equilibrium is appropriate.

It is easy to see that $\vec{\sigma}$ is a Nash equilibrium of a
standard game iff $\vec{\sigma}$ is a (generalized) Nash equilibrium
of the canonical representation of $\Gamma$ as a game with
awareness. Thus, our definition of generalized Nash equilibrium
generalizes the standard definition.

Consider the game with awareness shown in Figures \ref{fig:game1}
(taking this to be $\Gamma^m$), \ref{fig:A1game}, and
\ref{fig:A2game}. We have $\G_A=\{\Gamma^A,\Gamma^B\}$ and
$\G_B=\{\Gamma^m,\Gamma^B\}$. Taking $\dom(\sigma_{i,\Gamma'})$ to
denote the domain of the strategy $\sigma_{i,\Gamma'}$, we have
$$\begin{array}{l} \dom(\sigma_{A,\Gamma^A}) = \{(\Gamma^m,\<\,
\>),(\Gamma^A,
\<\mbox{unaware}\>),(\Gamma^A,\<\mbox{aware}\>)\},\\
\dom(\sigma_{B,\Gamma^m}) = \{(\Gamma^m,
\<\mbox{across}_A\>),(\Gamma^A,\<\mbox{aware, across}_A\>)\},\\
\dom(\sigma_{A,\Gamma^B}) = \{(\Gamma^B,\<\, \>)\},\mbox{ and }\\
\dom(\sigma_{B,\Gamma^B}) =
\{(\Gamma^A,\<\mbox{unaware,across}_A\>),(\Gamma^B,\<\mbox{across}_A\>)\}.
\end{array}$$
Each of these domains consists of a single generalized information
set. If $p<1/2$, then there exists a generalized Nash equilibrium where
$\sigma_{A,\Gamma^A}=$ across$_A$, $\sigma_{A,\Gamma^B}=$ down$_A$,
$\sigma_{B,\Gamma^m}=$ down$_B$, $\sigma_{B,\Gamma^B}=$ across$_B$.
Thus, in the modeler's game, $A$ plays across$_A$, $B$ plays
down$_B$, and the resulting payoff vector is $(2,3)$. On the other
hand, if $p > 1/2$, then there exists a generalized Nash equilibrium where
$\sigma_{A,\Gamma^A}=$ down$_A$, $\sigma_{A,\Gamma^B}=$ down$_A$,
$\sigma_{B,\Gamma^m}=$ down$_B$, $\sigma_{B,\Gamma^B}=$ across$_B$.
Thus, in the modeler's game, $A$ plays down$_A$, and the payoff
vector is $(1,1)$. Intuitively, even though both $A$ and $B$ are
aware of all the moves in the modeler's game, $A$ considers it
sufficiently likely that $B$ is not aware of
down$_B$, so $A$ plays down$_A$.
There exists another generalized Nash equilibrium where
$\sigma_{A,\Gamma^A}=$ down$_A$, $\sigma_{A,\Gamma^B}=$ down$_A$,
$\sigma_{B,\Gamma^m}=$ across$_B$,
and
$\sigma_{B,\Gamma^B}=$ across$_B$
that holds for any value of $p$.  Intuitively, $A$ believes $B$ will play
across$_B$ no matter what he ($B$) is aware of, and therefore plays
down$_A$; given that $A$ plays down$_A$, $B$ cannot improve by
playing down$_B$ even is he is aware of that move.%
\footnote{We did not discuss this latter equilibrium in the preliminary
version of this paper.}

We now show that every game with awareness has at least one
generalized Nash equilibrium.  We proceed as follows.  Given a game
$\Gamma^* = (\G, \Gamma^m, \F)$ with awareness, let $\nu$ be a
probability on $\G$ that assigns each game in $\G$ positive
probability.  (Here is where we use the fact that $\G$ is
countable.) We construct a standard
extensive
game $\Gamma^\nu$ by essentially ``gluing together'' all the games
$\Gamma' \in \G$,
except that we restrict to the histories in $\Gamma'$ that
can actually be played according to the players' awareness level.%
\fullv{
Formally, for each $\Gamma'\in \G$,
we restrict to the histories $\lfloor
H'\rfloor=\{h\in H':$ for every
prefix $h_1\cdot \<m\>$ of $h$, if $P'(h_1)=i \in N$ and
$\F(\Gamma',h_1)=(\Gamma'',I)$, then
for all $h_2\in I$, $h_2\cdot\<m\>\in \H''\}$.
}
\fullv{As we shall see, all the}\shortv{The}
components of $\Gamma^\nu$ are independent of $\nu$ except for
nature's initial move (as encoded by $f_c^\nu$). In $\Gamma^\nu$,
the set of players is $\{(i,\Gamma'):\Gamma'\in \G_i\}$. The game
tree of $\Gamma^\nu$ can be viewed as the union of
the pruned game trees of $\Gamma' \in \G$. The histories of
$\Gamma^\nu$ have the form $\<\Gamma'\>\cdot h$, where $\Gamma'
\in \G$ \fullv{and $h \in \lfloor\H^h\rfloor$.} \shortv{and $h$ is a
``pruned'' history in $\H^h$.} The move that a player or
nature makes at a history $\<\Gamma'\>\cdot h$ of $\Gamma^\nu$ is
the same as the move made at $h$ when viewed as a history of
$\Gamma'$. The only move in $\Gamma^\nu$ not determined by
$\Gamma^*$ is nature's initial move (at the history $\<\,\>$), where
nature chooses the game $\Gamma' \in \G$ with probability
$\nu(\Gamma')$.
\shortv{We leave the details of the construction of $\Gamma^{\nu}$
to the full paper.}

\fullv{Formally, let $\Gamma^\nu$ be a standard game such that
\begin{itemize}
\item $N^\nu = \{(i,\Gamma'):\Gamma'\in\G_i\}$;

\item $M^\nu = \G\union_{\Gamma'\in\G}\lfloor M'\rfloor$, where
$\lfloor M'\rfloor$ is the set of moves that occur in $\lfloor
H'\rfloor$;

\item $\H^\nu = \<\, \>\union\{\<\Gamma'\>\cdot h:\Gamma'\in\G, h\in
\lfloor H'\rfloor\}$;

\item $P^\nu(\<\, \>)=c$, and $$P^\nu(\<\Gamma^h\>\cdot h') =
\left\{ \begin{array}{ll}
(i,\Gamma^{h'}) &\mbox{if $P^h(h') = i \in N$ and}\\ &
\F(\Gamma^h,h')=(\Gamma^{h'}, \cdot),\\
c &\mbox{if $P^h(h') = c$;}\end{array} \right.$$

\item $f_c^\nu(\Gamma'|\<\, \>)= \nu(\Gamma')$ and
$f_c^\nu(\cdot|\<\Gamma^h\>\cdot h') = f_c^h(\cdot|h')$ if $P^h(h')
= c$;

\item $\I^\nu_{i,\Gamma'}$ is just the $\sim_i$ relation restricted to
histories $(\Gamma'',h)\in \H^{\nu}$ where $i$ moves and
$\F(\Gamma'',h)$ has the form $(\Gamma',\cdot)$;

\item $u_{i,\Gamma'}^\nu(\<\Gamma^h\>\cdot z)= \left\{
\begin{array}{ll}
u_i^h(z) &\mbox{if $\Gamma^h = \Gamma',$}\\
0 &\mbox{if $\Gamma^h \ne \Gamma'.$}\end{array} \right.$
\end{itemize}

}

\begin{theorem}
\label{THM:NASHEX} For all probability measures $\nu$ on $\G$
\begin{itemize}
\item[(a)] $\Gamma^{\nu}$ is a standard extensive game with perfect
recall; and

\item[(b)] if $\nu$ gives positive probability to all games in $\G$,
then $\vec{\sigma}$ is a Nash equilibrium of $\Gamma^\nu$ iff
$\vec{\sigma}'$ is a generalized Nash equilibrium of $\Gamma^*$,
where $\sigma_{i,\Gamma'}(\<\Gamma^h\>\cdot
h')=\sigma'_{i,\Gamma'}(\Gamma^h,h')$.
\end{itemize}
\end{theorem}

Although a Nash equilibrium does not necessarily exist in games with
infinitely many players, $\Gamma^\nu$ has three special properties:
(a) each player has only finitely many information sets, and (b) for
each player $(i,\Gamma')$, there exists a finite subset
$N(i,\Gamma')$ of $N^\nu$ such that $(i,\Gamma)$'s payoff in
$\Gamma^\nu$ depends only on the strategies of the players in
$N(i,\Gamma')$,
and (c) $\Gamma^\nu$ is a game with perfect recall.
This turns out to be enough to show that $\Gamma^\nu$ has at
least one Nash equilibrium. Thus, we get the following corollary to
Theorem~\ref{THM:NASHEX}.

\begin{corollary}\label{cor:Nashex}
Every game with awareness has a generalized Nash equilibrium.
\end{corollary}

\section{Modeling Awareness of Unawareness}
\label{sec:unawareness}

In this section, we describe how to extend our representation of
games with awareness to deal with awareness of unawareness. In an
augmented game that represents player $i$'s subjective view of the
game, we want to model the fact that $i$ may be aware of the fact
that $j$ can make moves at a history $h$ that $i$ is not aware of.
We do this by allowing $j$ to make a ``virtual move'' at history
$h$.
Histories that contain virtual moves are called {\em virtual
histories}. These virtual histories do not necessarily correspond to a
history in the underlying game $\Gamma$ (i.e., $i$ may falsely believe
that $j$ can make a move at
$h$ that he is unaware of), and even if a virtual history does
correspond to a history in $\Gamma$, the subgame that follows that
virtual history
may bear no relationship to the actual subgame that follows the
corresponding history in the underlying game
$\Gamma$. Intuitively, the virtual histories describe agent $i$'s
(possibly incomplete and possibly incorrect)
view of what would happen in the game if some move she is
unaware of is made by agent $j$. Player $j$ may have several
virtual moves available at history $h$, and may make virtual moves
at a number of histories in the augmented game.%
\footnote{In the preliminary version of the paper, we assumed that all
virtual moves were terminal moves.  This is appropriate if $i$ has no
idea at all of what will happen in the game after a virtual move is
made.  The greater generality we allow here is useful to model
situations where player $i$ has some partial understanding of the game.
For example, $i$ may know that he can move left after $j$'s virtual
move, no matter what that virtual move is.}
Note that agent $i$'s subjective game may include virtual moves for
$i$ himself; $i$ may believe that he will become aware of more moves
(and may take active steps to try and learn about these moves).

To handle awareness of unawareness, we consider a generalization of
the notion of augmented game.  We continue to refer to the
generalized notion as an augmented game, using ``augmented game
without awareness of unawareness'' to refer to the special case we
have focused on up to now. Formally, $\Gamma^+=(N^+, M^+, \H^+,
P^+,f_c^+,\{\I_i^+:i\in N^+\},\{u_i^+:i\in N^+\},\{\A_i^+ :i\in
N^+\})$ is an \emph{augmented game based on the (standard) finite
extensive game} $\Gamma=(N,M, \H,P,f_c,\{\I_i:i\in N\},\{u_i:i\in
N\})$ if it
satisfies conditions A1--A3, A6--A10 and A12 of augmented games, and
variants of A4, A5, and A8.%
\fullv{
\footnote{We could also relax A3 to allow some ``virtual
players''. We do not do that here for ease of exposition.}}
%
Before stating these variants we need to define formally the set of
virtual histories of $\Gamma^+$. The set of {\em virtual histories
$V^+$ of $\Gamma^+$} is defined by induction on the length of
histories as follows:
\begin{enumerate}
\item if $m\in H^+$, $m\in M^+-M$, and either $P^+(\emptyset)\in N^+$ or
$P^+(\emptyset)=c=P(\emptyset)$, then $m\in V^+$;
\item if $h\cdot\<m\>\in H^+$ and $h\in V^+$, then $h\cdot\<m\>\in V^+$;
\item if $h\cdot\<m\>\in H^+$, $m\in M^+-M$, $h\notin V^+$, and either
$P^+(h)\in N^+$ or $P^+(h)=c=P(\overline{h})$, then $m\in V^+$, where
if $h\notin V^+$, then $\overline{h}$ is the subsequence of $h$
consisting of all moves in $h$ that are also in $M$,
and if $h\in V^+$, then $\overline{h}=h$.
\end{enumerate}

We can now state the variants of A4, A5, and A8.
\begin{itemize}
\commentout{
\item[A4$'$.] If $P^+(h) \in N^+$, then $P^+(h)=P(\overline{h})$,
and $M^+_h\subseteq M_{\overline{h}} \union (M^+ -M)$. Moreover,
every history $h\cdot \<m\> \in \H^+$ with $m \in
M^+_h-M_{\overline{h}}$ is terminal. Intuitively, the moves in
$M^+_h - M_{\overline{h}}$ are virtual moves.
}
\item[A4$'$.] If $P^+(h) \in N^+$ and $h\notin V^+$, then
$P^+(h)=P(\overline{h})$
and $M^+_h\subseteq M_{\overline{h}} \union (M^+ -M)$.

\commentout{
\item[A5$'$.] If $P^+(h) = c$ and $M^+_h \inter M_{\overline{h}} \ne
\emptyset$, then $M^+_h\subseteq M_{\overline{h}} \union (M^+ -M)$.
Moreover,
if $m \in M^+h - M_{\overline{h}}$ and $M^+_h \inter
M_{\overline{h}} = \emptyset$, then $P(h \cdot \<m\>) \in N$; if $m
\in M^+h - M_{\overline{h}}$ and $M^+_h \inter M_{\overline{h}} \ne
\emptyset$, then $h \cdot \<m\>$ is terminal. Intuitively, if $M^+_h
\inter M^+_{\overline{h}} \ne \emptyset$, then all the moves in
$M^+_h-M_{\overline{h}}$ are virtual moves. }

\commentout{
\item[A5$'$.] If $P^+(h) = c$ and $M^+_h \inter M \ne
\emptyset$, then $M^+_h\subseteq M_{\overline{h}} \union (M^+
-M)$. Moreover, if $m \in M^+_h - M$ and $M^+_h \inter
M \ne \emptyset$, then $h \cdot \<m\>$ is terminal. Intuitively, if
$M^+_h \inter M \ne \emptyset$, then all the moves in $M^+_h-M$ are
virtual moves.}
\item[A5$'$.] If $P^+(h) = c$ and $h\notin V^+$, then either $P(\overline{h})=c$ and $M^+_h\subseteq M_{\overline{h}} \union (M^+
-M)$, or $P(\overline{h})\ne c$ and $M^+_h \cap M=\emptyset$.

\item[A8$'$.] If $h$ and $h'$ are in the same information set in
$\Gamma^+$ and $h,h'\notin V^+$, then $\overline{h}$ and $\overline{h}'$
are in the same information set in $\Gamma$.

\end{itemize}

A \emph{game with awareness of unawareness based on $\Gamma$} is
defined as a tuple $\Gamma^* = (\G, \Gamma^m, \F)$ just as before.
The modeler's extended game $\Gamma^m$ satisfies the same conditions
M1-M3 as before, and the mapping $\F$ satisfies C3--C5 and C7--C10
and the following variants of C1, C2, and C6:
\begin{itemize}
\item[C1$'$.] $\{\overline{h}:h\in H^h, h\notin V^h\}=a$.

\item[C2$'$.] If $h' \in \H^h$ and $P^h(h') = j$, then (a) $\A_j^h(h')
\subseteq a$, (b)  if $h' \notin V^h$, then $(M_{\overline{h}'}\inter \{m: \overline{h}'\cdot
\< m\>\in a\})\union (M^h_{h'}-M_{\overline{h}'})= M^h_{h'}$, and
(c) if $\F(\Gamma^h,h') = (\Gamma',I')$, then for all $h'' \in I'$,
we have $M'_{h''} \subseteq M^h_{h'}$.

\item[C6$'$.] If $h'\in I$ and $h,h'\notin V^h$, then $\overline{h}$ and
$\overline{h}'$ are in the same information set in $\Gamma$.
\end{itemize}
C1$'$ and C6$'$ have been weakened so that these restrictions only apply to
non-virtual histories of $\Gamma^h$.
Part (a) of C2$'$ is the same as the first part of C2; part (b)
implies that the
set of moves available to player $j$ at
a non-virtual history
$h'$ is the set of
moves that player $i$ is aware of that are available to $j$ at
$\overline{h}'$ in the underlying game together with some virtual moves.
It is not hard to check that in games without awareness of
unawareness, part (c)  follows from A4, C1, and C2, so it does not
need to be explicitly stated in C2.  However, now that A4 has been
weakened to A4$'$, we must mention it explicitly.

Note that $\Gamma^m$ is an augmented game with no awareness of
unawareness; there are no virtual moves, since the modeler is indeed
aware of all possible moves (and knows it). We can now define local
strategies, generalized strategy profiles, and generalized Nash
equilibrium just as we did for games with awareness. The same
technique as that used to show Corollary~\ref{cor:Nashex} can be
used to prove the following.

\begin{theorem}
Every game with awareness of unawareness has a generalized Nash
equilibrium.
\end{theorem}

\fullv{
\section{Modeling Lack of Common Knowledge}
\label{sec:CK}

Game theorists have long searched for good approaches to modeling
games where there is no common knowledge among players regarding the
game being played. Our approach is flexible enough to handle such
lack of common knowledge.
In this section, we discuss the changes needed to handle lack of
common knowledge. We remark that what we do here makes perfect sense
even in games where there is full awareness.


We can modify our model to accommodate four different aspects of
lack of common knowledge.

\begin{itemize}
\item {\em Lack of common knowledge regarding who moves.} We assumed
that every player
understands who moves in each history he is aware of. Although we
still need to require that every player knows when
it
is his turn to
move, we can handle the case where a player has false beliefs about
who moves after a history that is not in one of his information
sets.
For example, we are interested in modeling the case where player $i$
may be confused after some history $h$ as to whether player $j$ or
player $k$ moves, but in both cases $i$ still believes that the same
moves are available. That is, player $i$ knows what could happen
next, but he does not know who is going to do it.
(Later we model
uncertainty not only regarding who moves but also regarding what the
move is.)

To explain the necessary modifications, we need one more definition.
Let $\G_{m,i}$ be the smallest subset of $\G$ such that if either
$\Gamma^+ = \Gamma^m$ or $\Gamma^+ \in \G_{m,i}$, $h\in H^+$,
$P^+(h)=i$, and $\F(\Gamma^+,h)=(\Gamma',\cdot)$, then
$\Gamma'\in\G_{m,i}$. Intuitively, $\G_{m,i}$ consists of all games
player $i$ considers possible, or considers possible that he
considers possible, and so on, at some history of the modeler's
game.

We can model lack of common knowledge about who moves by replacing
A4 by
\begin{itemize}
\item[A4$''$.] If $P^+(h)=i\in N^+$, then $M^+_{h}\subseteq
M_{\overline{h}}$.
\end{itemize}

Thus, we no longer require that the player who moves at history $h$ is
necessarily the one who moves at $\overline{h}$.
However, we do make this requirement for the
modeler's game,
since the modeler is assumed to understand the underlying game.
Thus,
we must add a requirement M4 for the modeler's game that is
identical to A4 except that we replace $\Gamma^+$ by $\Gamma^m$.

Player $i$ must also understand that he moves at a history $h$ iff
he moves at $\overline{h}$ for games in $\G_{m,i}$.
\begin{itemize}
\item[C11.] If $\Gamma^+\in \G$, $h \in
\H^+$, $P^+(h) = i$, $\A_i^+(h) = a$, $\F(\Gamma^+,h) =
(\Gamma^h,I)$, and $h'\in \H^h$, then
if $\Gamma^h\in \G_{m,i}$ and $P^h(h')=i$, then $P(\overline{h}')=i$. Conversely, if
$P(\overline{h}')=i$, then there exists a prefix or suffix $h''$ of
$h'$ such that $\overline{h}''=\overline{h}'$ and $P^h(h'')=i$.
\end{itemize}

\commentout{ There is an additional requirement.  $\Gamma^+$ is
either the modeler's game or some players beliefs about \ldots\ some
player's beliefs about what the actual game is.  We require a
certain consistency in these beliefs.  If $P^+(h') = i$ then
whoever's view of the game is represented by $\Gamma^+$, that player
believes that the signals in $h'$ will result in $i$'s moving.  For
consistency, $i$ should believe that it is $i$'s move in $h'$. Thus,
we require the following condition:
\begin{itemize}
\item[C11.]
If $h'$ is a history of both $\Gamma^+$ and $\Gamma^h$,
then $P^+(h')=i$ iff $P^h(h')=i$.
\end{itemize}
}


We also need to make modifications to A5.
Since we want to allow a player to have false beliefs about when nature
moves, we replace A5 with
\begin{itemize}
\item[A5$''$.] If $P^+(h) = c$, then either
$M^+_h \subseteq M_{\overline{h}}$, or $M^+_h \inter M = \emptyset$.
\end{itemize}
As before, the moves in $M^+_h$ in the case where $M^+_h \inter M=
\emptyset$ intuitively capture
uncertainty regarding a player's awareness level. But now it may be
the case that a player $i$ falsely believes that nature moves after
history $\overline{h}$ in the underlying game.
Just as with A4, we must add a condition M5 to the modeler's game that
is identical to A5, except that $\Gamma^+$ is replaced by $\Gamma^m$.

\item {\em Lack of common knowledge about the information sets.} We
assumed that
every player understand the signals every other player receives in
every history he is aware of. We can weaken this assumption by
allowing a player to have false beliefs about the signals received
by other players, or equivalently, by allowing a player to have
false beliefs about the information sets of other players.

We can model lack of common knowledge about the information sets by
removing conditions A8--A10.  Again, because we assume that the
modeler understands the information sets, we would add
analogues of A8--A10 to the conditions on the modeler's game
(replacing $\Gamma^+$ by $\Gamma^m$, of course).  Similarly, we
would require analogues of A8 and A9 to hold in the ``C-list'' of
conditions for games $\Gamma^h \in \G_{m,i}$, and we weaken C6 so
that it also holds only for $\Gamma^h \in \G_{m,i}$.
We must also add an analogue of A10 to the ``C-list'' for games
$\Gamma^h\in \G_{m,i}$ for histories $h'$ and $h''$ in an
$i$-information set.
\commentout{ We also require a condition analogous to C11.
\begin{itemize}
\item[C12.] If $\Gamma^+\in \G$, $h \in
\H^+$, $P^+(h) = i$, $\A_i^+(h) = a$ and $\F(\Gamma^+,h) =
(\Gamma^h,I)$, $h'$ and $h''$ are histories in both $\Gamma^+$ and
$\Gamma^h$, then $h'$ and $h''$ are in the same $i$-information set
in $\Gamma^+$ iff $h'$ and $h''$ are in the same $i$-information set
in $\Gamma^h$.
\end{itemize}
Although the player $j$ (or the modeler) whose view of the game is
$\Gamma^+$  may have false beliefs about player $i$'s information
sets, player $j$ knows that $i$ knows his information sets. C12
captures that intuition. }

\item {\em Lack of common knowledge about payoffs.} We assumed that payoffs
depended only on moves of the underlying game and that they were
common knowledge among players.
By dropping condition A12, we remove both of these assumptions.
If we want to require that payoffs depend only on the underlying
game, but still want to allow players to have false beliefs about the
utilities, we would
add an analogue of A12 in the modeler's game and
use the following weakening of A12:
\begin{itemize}
\item[A12$'$.] If $\Gamma^+\in \G$, $z,z' \in
Z^+$, and $\overline{z}=\overline{z}'$, then for all $i\in N^+$,
$u_i^+(z)=u_i^+(z')$.
\end{itemize}
Although the player $j$ whose view of the game is $\Gamma^+$  may
have false beliefs about the payoffs, player $j$ knows that the
payoffs depend only on the moves made in the underlying game. A12$'$
captures that intuition.

\commentout{
There are two ways of modeling lack of common knowledge about
payoffs depending on whether we interpret payoffs of the underlying
game as objective (e.g., dollars) or subjective (e.g., utilies). In
both cases we remove condition A12.

If we view payoffs of the underlying game as subjective, then if
$\Gamma^+$ is an augmented game considered possible by some player
at some situation the only meaningful payoffs of this game are those
of player $i$. Since the payoff of a player may depend on the game
he considers possible and in $\Gamma^+$ players other than $i$ may
consider different games possible, their payoffs in $\Gamma^+$ are
meaningless and irrelevant. Note that in our definition of
generalized Nash equilibrium those payoffs were not used. There is
no restriction on what the payoffs in the augmented games should be
in this case.

If we view payoffs as objective, then we need an analogue of A12 for
the modeler's game. We also require the following condition:
\begin{itemize}
\item[A12$'$.] If $\Gamma^+\in \G$, $z,z' \in
Z^+$, $\overline{z}=\overline{z}'$, then for all $i\in N^+$,
$u_i^+(z)=u_i^+(z')$.
\end{itemize}
Although the player $j$ whose view of the game is $\Gamma^+$  may
have false beliefs about the payoffs, player $j$ knows that the
payoffs depend only on the moves made in the underlying game. A12$'$
captures that intuition.
}

\item {\em Lack of common knowledge of the underlying game.} We assumed
players have common knowledge about the structure of the underlying
game. Our framework
can model a situation where each player has a completely different
conception of what game is actually being played, which may have
very little relationship to the actual underlying game (although we
still assume that the modeler's game corresponds to the actual
game).
The key idea is to drop the assumption that all augmented games are
based on the same game $\Gamma$.

To formalize this intuition, we modify A2 so that the $\A_i^+$ function
does not necessarily map histories of an augmented game to histories of
the same game $\Gamma$.
Rather, $\A^+(h)$ is the set of histories of some game $\Gamma(h)$ that,
intuitively, $i$ considers to be the true underlying game.
Thus, if $h$ and $h'$ are two histories in $\Gamma^+$, then $\A_i^+(h)$
and $\A_i^+(h')$ may be histories
in two completely different games.
\commentout{ We retain A1 and A6, and, as before, replace A4 and A5
by A4$''$ and A5$''$, respectively, and drop A3 and A7--12.}
Since $\Gamma(h)$
is viewed as $i$'s subjective view of the true underlying game,
we assume
that
he understands it perfectly. Thus, we retain A1, A6
and A8--A12 and replace conditions A3--A5 by M1--M3
(where the set of players is the set of players in $\Gamma(h)$ and the
projection function maps a history $h$ to a history $\overline{h}$ in
$\Gamma(h)$).
With regard to A7, note that, even if a player intuitively has perfect
recall, he
may realize in the future that he does not consider possible
a history he considered possible in the past.

In the definition of games with awareness, we allow
$\G$ to contain augmented games based on standard games different
from the game on which $\Gamma^m$, the modeler's game, is based.
We continue to require conditions C1, C3--C5, C7--C9, and C11,
%
but we weaken C2.
In C2 we required that a player $i$ cannot consider possible
a game $\Gamma^h$ where one of the players $j$ moving in $\Gamma^h$
is aware of more runs than $i$ is. In this setting, we allow $i$ to
consider possible a game $\Gamma^h$ where one of the players $j$
moving in $\Gamma^h$
believes (falsely, from $i$'s point of view) that some runs are possible
that $i$ does not consider possible.  However, we require that
the set of moves that $i$ believes that $j$ believes are available
to him while moving at history $h'$ in
$\Gamma^h$ is a subset of the moves $i$ believes are
available to $j$ while moving at $h'$.
We thus replace C2 by the following condition C2$''$, which is the
analogue of parts (b) and (c) of C2$'$.
\begin{itemize}
\item[C2$''$.] If $h' \in \H^{h}$, $P^{h}(h') = j$, and $\F(\Gamma^h,h')=(\Gamma',I')$, then for all $h''\in I'$,
$M'_{h''}\subseteq M^h_{h'}$,
and $M_{\overline{h}'}\inter \{m: \overline{h}'\cdot\< m\>\in a\}=
M^{h}_{h'}$.
\end{itemize}

Since we allow players to have false beliefs about information
sets, we drop conditions C6 and C10. However, since we have dropped
A7 and
weakened C2, we now need the following condition, which requires that
if a player considers possible a set of histories of the underlying
game, then he cannot believe that in the future he will consider
possible a different set of histories.\footnote{Note that this does
not rule out a situation where a player
$i$ realizes at history $h'$ that his view of
the game will change at a future history $h''$ when he receives some
additional information. If this
is the case, then this should already be described in the set of
histories that $i$ considers possible at $h'$.}
\begin{itemize}
\item[C12.] If $h' \in I$, $h''\in \H^{h}$, $P^{h}(h'') = i$, and
$h''$ is a suffix of $h'$, then $\A_i^h(h')=\A_i^h(h'')$.
\end{itemize}
It is easy to see what C12 follows from A7, C2, and C4, which is why we
did not list it explicitly earlier.

This approach of
allowing the augmented games in $\G$ to be based on different
underlying games actually
subsumes our earlier approach and allows us to capture lack of
common knowledge about who moves, what the information sets are, and
what the payoffs are.
For example, note that despite the fact that we have replaced A3-5 by M1-3,
we can also model games with awareness using this approach
by taking the game $\Gamma(h)$ to be the game consisting only of the
runs of $\Gamma$ that are in $\A^+(h)$.
(Of course, if we do that, we need to reinstate A7 and replace C2$''$
with C2.)
To capture lack of common knowledge about who moves, we take $\Gamma(h)$
to be identical to $\Gamma$ except that different agents may move at a
given information set.
Similarly, we can model lack of common knowledge about
what the information sets and what the payoffs are by restricting
$\Gamma(h)$ appropriately.

\commentout{ Most of the conditions A1--11 now disappear.
Specifically, we retain A1 and A6, modify A2 so that $A_i^+$ maps
histories in $H_i^+$ to sets of histories of $\Gamma$; now $A_i^+$
simply maps histories $h$ of an augmented game $\Gamma^+$ to
histories of some standard game $\Gamma(h)$ (not necessarily
$\Gamma$).  All the other conditions disappear.  On the other hand,
we do need conditions M1--5 in the modeler's game (recall that M4
and M5 are analogues of A4 and A5), since the modeler is presumed to
understand the true underlying game.

Conditions C1--11 remain, except that we must weaken C2 and add an
analogue of A5.

\commentout{
Formally, we model this by defining a {\em
subjective game based on $\Gamma$} to be a tuple
$\Gamma^{*}=(\G,\Gamma^m,\F)$, where
\begin{itemize}

\item $\G$ is a countable set of augmented games based on games in
$\J$, including $\Gamma^m$ that is an augmented game based on
$\Gamma$.

\item $\F$ maps an augmented game $\Gamma_1^+\in \G$ and a history
$h$ in $\Gamma_1^+$ such that $P^+(h)=i$ to a pair $(\Gamma_2^h,I)$,
where $\Gamma_2^h\in \G$ and $I$ is an $i$-information set in game
$\Gamma_2^h$.
\end{itemize}

The definition of augmented games is exactly the same, only instead
of interpreting $\A_i^+$ to consist of the set of runs an agent is
aware of, we interpret it as the set of runs an agent considers
possible. The set of constraints in the modeler's game $\Gamma^m$ is
exactly as in the case of games with awareness. The conditions in
the other augmented games must be slightly modified. Since we now
have more than one underlying game, when referring to components of
an augmented game $\Gamma^+$ instead of using superscript $+$ we use
$\Gamma^+$. Let $\Gamma_1^+\in \G$, $h\in H^{\Gamma_1^+}$,
$P^{\Gamma_1^+}(h)=i$, $A_i^{\Gamma_1^+}(h)=a$, and
$\F(\Gamma_1^+,h)=(\Gamma_2^h,I)$. We then require the following
conditions:
\begin{itemize}
\item Conditions S1, S3-S9 that are analogues of C1, C3-C9 using the
new notation. For example, S8 is

\begin{itemize}
\item[S8.] For all histories $h'\in I$, there exists a prefix $h'_1$ of
$h'$ such that $P^{\Gamma_2^h}(h'_1)=i$ and
$\F(\Gamma_2^h,h'_1)=(\Gamma_3',I')$ iff there exists a prefix $h_1$
of $h$ such that $P^{\Gamma_1^+}(h_1)=i$ and
$\F(\Gamma_1^+,h_1)=(\Gamma_3',I')$. Moreover, $h'_1\cdot \<m\>$ is
a prefix of $h'$ iff $h_1\cdot \<m\>$ is a prefix of $h$.
\end{itemize}

\item Condition S2 that is a weaker version of C2.}

In C2 we required that a player $i$ cannot consider possible a game
$\Gamma^h$ where one of the players $j$ moving in $\Gamma^h$ is
aware of more runs than $i$ is. In this setting, we allow $i$ to
consider possible a game $\Gamma^h$ where one of the players $j$
moving in $\Gamma^h$ falsely believes in more runs than $i$
believes. However, we require that the set of moves that $i$
believes that $j$ believes are available to him while moving at
history $h'$ in
$\Gamma^h$ is a nonempty subset of the moves $i$ believes are
available to $j$ while moving at $h'$.
We thus replace C2 by the following condition C2$''$, which is the
analogue of part (c) of C2$'$.
\begin{itemize}
\item[C2$''$.] If $h' \in \H^{h}$, $P^{h}(h') = j$, and
$\F(\Gamma^h,h')=(\Gamma',I')$, then for all $h''\in I'$,
$M'_{h''}\subseteq M^h_{h'}$, and
$M^{\Gamma(h)}_{\overline{h}'}\inter \{m: \overline{h}'\cdot\<
m\>\in a\}= M^{h}_{h'}$.
\end{itemize}


\commentout{
\begin{itemize}
\item[S1.]$\{\overline{h}: h \in \H^{\Gamma_2^h}\} = a$.

\item[S2.] If $h' \in \H^{\Gamma_2^h}$, $P^{\Gamma_2^h}(h') = j$ and
$\F(\Gamma_2^h,h')=(\Gamma_3',I')$, then for all $h''\in I'$
$\emptyset\ne M^{\Gamma_3'}_{h''}\subseteq M^{\Gamma_2^h}_{h'}$ and
$M^{\Gamma_2}_{\overline{h}'}\inter \{m: \overline{h}'\cdot\< m\>\in
a\}= M^{\Gamma_2^h}_{h'}$.

\item[S3.] If $h'$ and $h$ are in the same information set in
$\Gamma_1^+$ and $\overline{h}' \in a$, then there exists $h'' \in
I$ such that $\overline{h}''=\overline{h}'$.

\item[S4.] If $h'\in I$, then $\A_i^{\Gamma_2^h}(h')=a$ and
$\F(\Gamma_2^h,h')=(\Gamma_2^h,I)$.

\item[S5.] If $h' \in \H^{\Gamma_1^+}$, $P^{\Gamma_1^+}(h') = i$, $\A_i^{\Gamma_1^+}(h')=a$,
then if $h$ and $h'$ are in the same information set of
$\Gamma_1^+$, then $\F(\Gamma_1^+,h') = (\Gamma_2^h,I)$, while if
$h$ is a prefix or a suffix of $h'$, then $\F(\Gamma_1^+,h') =
(\Gamma_2^h,I')$ for some $i$-information set $I'$.

\item[S6.]
If $h' \in I$, then $\overline{h}$ and $\overline{h}'$ are in the
same information set in $\Gamma_2$;

\item[S7.] If $\Gamma_2^h = \Gamma_1^+$, then $h \in I$.

\item[S8.] For all histories $h'\in I$, there exists a prefix $h'_1$ of
$h'$ such that $P^{\Gamma_2^h}(h'_1)=i$ and
$\F(\Gamma_2^h,h'_1)=(\Gamma_3',I')$ iff there exists a prefix $h_1$
of $h$ such that $P^{\Gamma_1^+}(h_1)=i$ and
$\F(\Gamma_1^+,h_1)=(\Gamma_3',I')$. Moreover, $h'_1\cdot \<m\>$ is
a prefix of $h'$ iff $h_1\cdot \<m\>$ is a prefix of $h$.

\item[S9.] There exists
a history $h'\in I$ such that for every prefix $h''\cdot \<m\>$ of
$h'$, if $P^{\Gamma_2^h}(h'')=j \in N^{\Gamma_2^h}$ and
$\F(\Gamma_2^h,h'')=(\Gamma_3',I')$, then for all $h_1\in I'$,
$h_1\cdot \<m\>\in \H{\Gamma_3'}$.

\end{itemize}
}

\commentout{
It is not difficult to see that games with awareness are a special
case of subjective games where $\J=\{\Gamma\}$. We also can use this
framework to model the cases of lack of common knowledge regarding
who moves, about payoffs and information sets described before by
restricting the set $\J$ appropriately. In the same spirit of
Section~\ref{sec:unawareness}, it is possible to extended subjective
games to deal with the case of beliefs about lack of beliefs; we
leave details for the reader.
}

}
\end{itemize}

Despite all the changes to the conditions,
the definitions of local strategies and generalized Nash
equilibrium, and the theorems and their proofs remain unchanged.
Thus, our techniques can deal with highly subjective games as well as
awareness.

\commentout{
Of course, in  many cases of interest, extra properties will hold.
For example, consider a large
geographically-dispersed game where agents interact only with nearby
neighbors.  In such a game, an agent may be unaware of exactly who is
playing the game
(although she may realize that there are other
players besides her neighbors, and even realize that the moves
made by distant players may have an indirect effect on her).  To
model such a situation, we may want to have virtual moves after
which the game does not end, and to allow agents to be aware of
subsequences of histories in the underlying game.
}

\section{Related Work}
\label{sec:liter}

There have been a number of models for unawareness in the literature
(see, for example, \cite{FH,HMS03,MR94,MR99,DLR98}). Halpern
\citeyear{Hal34} and Halpern and R\^ego \citeyear{HR05} showed that
in a precise sense all those models are special cases of Fagin and
Halpern's \cite{FH} approach where they modeled awareness
syntactically by introducing a new modal operator for it.
Halpern and R\^ego \cite{HR05b} extended Fagin and Halpern's logic
of awareness to deal with knowledge of unawareness.
All of these papers focused on logic, and did not
analyze the impact of unawareness in a strategic setting.

Feinberg's \citeyear{Feinberg04,Feinberg05} work is most similar
work to ours.
We discussed the high-level difference between our work and that of
Feinberg in the introduction. Here we focus on some of the more detailed
differences:
\begin{itemize}
\item Feinberg does not model games semantically.
He encodes all the information in the $\F$ function syntactically, by
describing each player's awareness level and iterated nested awareness levels
(e.g., what player $1$ is aware that player $2$ is aware that player $3$
is aware of).

\commentout{
\item In dealing with normal-form games, Feinberg~\citeyear{Feinberg05}
imposes two
restrictions: (a) if player $i$ is aware that player $j$ is aware of
move $m$, then player $j$ must indeed be aware of move $m$, and (b)
for every player $i$, $i$ is aware of all
the moves he can make.
Moreover, there is higher-order awareness of these facts.
In our framework,
(a)  amounts to assuming that if $P^+(h) = j$, $\F(\Gamma^+,h) =
(\Gamma^h,I)$, $h'$ is a history in $\Gamma^h$ such that $P(h') =
j$, then for all histories $h''$ in $\Gamma^+$ such that $P(h'') =
j$, $A_j^h(h') \subseteq A_j^+(h'')$. It easily follows from this
condition and C4 that $A_j^+(h'')$ is the same at all histories
$h''$ in $\Gamma^+$.  Applying C4 again, we can conclude that $j$'s
awareness level is the same at all histories in all games in $\G$.
Thus, there is no uncertainty about awareness. Feinberg's second
restriction amounts to strengthening the subset relations in A4 to
equalities.  That is, if $\Gamma^+$ is an augmented game, then $\{m:
\overline{h}\cdot \<m\> \in A_i^+(h)\} = M^+_h  = M_{\overline{h}}$.
In the presence of the first restriction, the second restriction can be
made without loss of generality.  However, the first restriction is
quite severe.
Even in the normal-form setting, it prevents us
from capturing many games of interest.

}

\item
In dealing with extensive games, Feinberg~\citeyear{Feinberg04}
assumes that the runs that a player is aware of completely determine
what game he believes he is playing. It cannot be the case
that there are two distinct ``identities'' of a player that have the same
awareness level.
\commentout{Thus, for example, if player 1 and player 2 are both
contemplating an instance of player 3 who is aware only of two moves
and all three players, they will necessarily view associate the same
strategy with player 3.  It also cannot be the case that the game
that a player believes he is playing depends on the order in which
he became aware of moves.  While this assumption does not affect the
question of whether there exists a generalized Nash equilibrium, it
does restrict the set of possible generalized Nash equilibria.}
As we discussed in Section~\ref{sec:unaware}, this assumption limits
the applicability of the model.

\commentout{
\item In dealing with normal-form games, Feinberg~\citeyear{Feinberg05}
assumes that every player is aware of his own actions and that there
is higher-order awareness of this fact.   This is not unreasonable for
games where there is a single move, but is inappropriate for
extensive-form games.  We do not assume it here, nor does
Feinberg~\citeyear{Feinberg04} when he deals with extensive-form games.
}

\item Feinberg
assumes that if player $i$ is aware of player $j$, then
$i$ must be aware of some move of player $j$. We do not require such
a condition since
the analogous condition is not typically assumed in standard extensive games.
For example, in a standard extensive game, a player may get a payoff
even though there is no node where he can move.
%
But it is trivial to add this requirement (as it would be trivial to
drop in Feinberg's framework), and making it has no impact on the
results.

\item Feinberg~\citeyear{Feinberg05}
defines payoffs for player $i$ by
using what he calls ``default actions'' for players that $i$ is
unaware of. He says that this default action will be context
dependent. We do not have such default actions in our setting; the
payoff of a player in our framework is independent of the payoff of the players
he is unaware of.
The assumption of a default action seems somewhat problematic to us; it
is not clear what the default move should be in general.
Moreover, if two different players are unaware of player $j$,
it is not clear why (or whether) they should assume the same default
action.

\item In dealing with extensive games, Feinberg~\citeyear{Feinberg04} defines
moves of nature
by conditioning on the set of moves of nature the player is aware
of. In our framework,
this would amount to the
following requirement:
\begin{enumerate}
\item[C13] If $\Gamma^+\in \G$, $h \in
\H^+$, $P^+(h) = i$, $\A_i^+(h) = a$, $\F(\Gamma^+,h) =
(\Gamma^h,I)$, $h'\in H^h$, $P^h(h')=c$, and $M^h_{h'} \inter
M_{\overline{h}'} \ne \emptyset$, then $\mbox{$f^h_c( m \mid h')$} =
\frac{f_c(m \mid \overline{h}')}{f_c(M^h_{h'}\mid \overline{h}')}$
for every $m\in M^h_{h'}$ and $f^h_c( m \mid h')=0$ if $m\notin
M^h_{h'}$.
\end{enumerate}
As Feinberg did, for that condition to be well defined we require
that $f_c(m\mid \overline{h})\ne 0$ for all $m\in M_{\overline{h}}$
and histories $h$.
As we discussed in Section~\ref{sec:unaware}, while we believe such
a requirement makes sense if nature's move is interpreted
objectively, it does not make sense in general so we do not assume
this in every augmented game.

\end{itemize}

Sadzik~\citeyear{Sadzik05} considers a logic of awareness, knowledge,
and probability based on that of
Heifetz, Meier, and Schipper \citeyear{HMS03}, and uses it to give a
definition of Bayesian equilibrium in normal-form games with awareness.
Heifetz, Meier and Schipper~\citeyear{HMS06} also consider a
generalized state-space model with interactive unawareness and
probabilistic beliefs and give a definition of Bayesian equilibrium in
normal-form games, without assuming Feinberg's restriction.
Li \citeyear{LI06b} has also provided a model of unawareness in
extensive games, based on her earlier work on modeling
unawareness \cite{LI06,LI06a}. Although her representation of a game
with unawareness is quite similar to ours, her notion of generalized
Nash equilibrium is different from ours.   Just as we do, she
requires that every player $i$ make a best response with respect to his
beliefs regarding
other player's strategies in the game $\Gamma^i$ that $i$ considers
possible. However, unlike us, she requires that these beliefs
satisfy a consistency requirement that implies, for example, that if
a player $i$ is aware of the same set of moves for him at
both information set $I_1$ in game $\Gamma_1$ and information set $I_2$
in $\Gamma_2$,
and these information sets correspond to the same
information set in the underlying game $\Gamma$, then the local
strategies $\sigma_{i,\Gamma_1}$ and $\sigma_{i,\Gamma_2}$ must
agree at these information sets; that is, $\sigma_{i,\Gamma_1}(I_1) =
\sigma_{i,\Gamma_2}(I_2)$.
\commentout{
by
projecting the strategies used by those players in the game they
consider possible onto the game $\Gamma^i$.}

Ozbay \citeyear{Ozbay06} proposes a model for games with uncertainty
where players may have different awareness levels regarding a move
of nature. He assumes that one of the players is
fully aware, and
can tell the other player about these moves before the second player
moves. Although our model can easily capture
this setting, what is interesting about Ozbay's approach is that
the second player's beliefs about the probability of these revealed
moves of are
formed as part of the equilibrium definition. Filiz
\citeyear{Filiz06} uses Ozbay's model in the context of
incomplete contracts in the presence of unforseen contingencies. In
this setting, the insurer is assumed to be fully aware of the
contingencies, and to decide strategically which contingencies to
include in a contract, while the insuree may not be aware of
all possible contingencies.

Finally, we remark that our notion of a game with awareness as
consisting of the modeler's game together with description of which
game each agent thinks is the actual game at each history has much
in common with the intuition behind Gal and Pfeffer's
\citeyear{GP05} notion
of a
\emph{Network of Influence Diagrams (NID)}.
Formally,
NIDs are a
graphical language for representing uncertainty over decision-making
models.
A node in a NID (called a \emph{block} by Gal and Pfeffer)
represents an agent's subjective belief about the underlying game and
what the strategies used by agents depend on.
Each node (game) in a NID is associated with a \emph{multiagent
influence diagram} \cite{KM01a}
(MAID), which is a compact representation of a game.
A NID has directed edges between nodes labeled by pairs of the form $(i,H)$,
where $i$ is an agent and (in our language) $H$ is a set of histories.
Intuitively, if there an edges from a node (game) $\Gamma$ to a node
$\Gamma'$ in a NID labeled
by a pair $(i,H)$, then $H$ is a set a set of histories in $\Gamma$,
there is an agent $j$ that moves at all the
histories in $H$, and in game $\Gamma$, $i$ believes that $j$ believes
that $\Gamma'$ is the true game when moving at a history $h \in H$.

Although Gal and Pfeffer do not try to handle notions of awareness with
NIDs, it seems possible to extend them to handle awareness. To do
this appropriately,
consistency requirements similar to
C1--C10 will need to be imposed.

}

\section{Conclusion}
\label{sec:conc}

We have generalized the representation of games to take into account
agents who may not be aware of all the moves
\fullv{ or all the other agents,}
\shortv{or agents,}
but may be aware of their lack of awareness.
\fullv{Moreover,}
\shortv{As we show in the full paper,}
our representation is also flexible enough to deal with
subjective games when there is lack of common knowledge about the
game, even if awareness is not an issue.
We have also shown how to define strategies and Nash equilibrium in
such settings. These generalizations greatly increase the
applicability of game-theoretic notions in multiagent systems.
In large games involving many agents, agents will almost
certainly not be aware of all agents and may well not be aware of
all the moves that agents can make. Moreover, \fullv{as we suggested in the
introduction, even in well-understood games like chess,} by giving
awareness a more computational interpretation, we can provide a more
realistic model of \fullv{the game} \shortv{a game like chess.}
\fullv{from the agents' perspective.}
We remark that although we focus on generalizing extensive-form
games, our framework is able to deal with normal-form games as well,
since we can view normal-form games as a special case of
extensive-form games.

There is clearly much more to be done to understand the role of
awareness (and lack of awareness) in multiagent systems.  We list
some of the many issues here:
\begin{itemize}
\fullv{
\item We have assumed perfect recall here.  But in long games,
it seems more reasonable to assume that agents do not have perfect
recall.  In a long chess game, typical players certainly do not
remember all the moves that have been played and the order in which
they were played.  It is well known that even in single-agent games,
considering agents with imperfect recall leads to a number of
subtleties (c.f.~\cite{Hal15,PR97}).   We suspect that yet more
subtleties will arise when combining imperfect recall with lack of
awareness.}

\item
In a Nash equilibrium of an extensive-form game, it may be the case
that the move made at an information set is not necessarily a best
response if that information set is not reached.
\fullv{For example, in the game described in Figure~\ref{fig:game1},
even if both players have common knowledge of the game, the profile
where $A$ moves down and $B$ moves across is a Nash equilibrium.
Nevertheless moving down is not a best response for $B$ if $B$ is
actually called upon to play. The only reason that this is a Nash
equilibrium is that $B$ does not in fact play. \emph{Sequential
equilibrium} \cite{KW82} is a solution concept that is arguably more
appropriate for an extensive-form game; it refines Nash equilibrium
(in the sense that every sequential equilibrium is a Nash
equilibrium) and does not allow solutions such as (down$_A$,
across$_B$).}
\shortv{There are refinements of Nash equilibrium that are arguably
more appropriate for an extensive-form game \cite{OR94}.} Our
representation of games with awareness (of unawareness) allows for relatively
straightforward generalizations of such refinements of Nash
equilibrium.
However, there are subtleties involved in showing that
generalized versions of these refinements always exist.
\fullv{For example, we no longer have a one-to-one correspondence
between the generalized sequential equilibria of the game $\Gamma^*$
and the sequential equilibria of the corresponding standard game
$\Gamma^{\nu}$.
Nevertheless, we
believe that we should be
able to use a more refined construction to show that a generalized
sequential equilibrium exists in every game with awareness.}

\item We have analyzed situations where agents may be unaware of some
moves in the underlying game%
\fullv{,}\shortv{ and}
may be aware of their unawareness%
\shortv{.}\fullv{, and may have completely false beliefs about the
underlying game.} Of course, there are other cases
of interest where additional properties may hold. For example, consider
a large geographically-dispersed game where agents interact only
with nearby neighbors.  In such a game, an agent may be unaware of
exactly who is playing the game (although she may realize that there
are other agents besides her neighbors, and even realize that the
moves made by distant agents may have an indirect effect on her). To
model such a situation, we may want to have virtual moves after
which the game does not end, and to allow agents to be aware of
subsequences of histories in the underlying game. We suspect that
\fullv{a straightforward extension}
\shortv{extensions} of the ideas in this paper can deal with such
situations, but we
have not worked out the details.

\item There has been a great deal of work on computing Nash equilibria.
As we have shown, a generalized Nash equilibrium of a game with
awareness is a Nash equilibrium of a standard game.  However, this
standard game can be rather large.  Are there efficient computational
techniques for computing generalized Nash equilibrium in interesting
special cases?

\item If there is little shared knowledge regarding the underlying game,
the set $\G$ of augmented games can be quite large, or even
infinite. Is it important to consider all the iterated levels of
unawareness encoded in $\G$?  Halpern and Moses
\fullv{\citeyear{HM1}} \shortv{\citeyear{HM90}} showed that, in
analyzing coordinated attack, no finite level of knowledge suffices;
common knowledge is needed for coordination. Stopping at any finite
level has major implications.
\fullv{Rubinstein \citeyear{Rub89} considered a variant of the
coordinated attack problem with probabilities,  and again showed
that no finite level suffices (and significant qualitative
differences arise if only a finite part of hierarchy of knowledge is
considered).}
On the other hand, Weinstein and Yildiz \citeyear{WY03} provide a
condition under which the effect of players' $k$th order beliefs is
exponentially decreasing in $k$. While we strongly suspect that
there are games in which higher-order unawareness will be quite
relevant,%
\fullv{ just as with the Weinstein-Yildiz result,} there may be
conditions under which higher-order awareness becomes less
important,
and a simpler representation may suffice.
Moreover, it may be possible to use NIDs to provide a more compact
representation of games of awareness in many cases of interest (just as
Bayesian networks provide a compact representation of probability
distributions in many cases of interest), leading to more efficient
techniques for computing generalized Nash equilibrium.
\end{itemize}

\fullv{We hope to explore some of these issues in forthcoming work.}

\commentout{
\section{Acknowledgments}
We thank Kobi Gal, Aviad  Heifetz, Bart Lipman, Avi Pfeffer, and Ilya
Segal for useful
This work was supported in part by NSF under grants
CTC-0208535 and ITR-0325453, by ONR under grants N00014-00-1-03-41
and N00014-01-10-511, and by the DoD Multidisciplinary University
Research Initiative (MURI) program administered by the ONR under
grant N00014-01-1-0795. The second author was also supported in part
by a scholarship from the Brazilian Government through the Conselho
Nacional de Desenvolvimento Cient\'ifico e Tecnol\'ogico (CNPq).}

\bibliographystyle{abbrv}
\bibliography{z,joe}

\appendix
\section{Proofs}


\othm{THM:NASHEX} For all probability measures $\nu$ on $\G$
\begin{itemize}
\item[(a)] $\Gamma^{\nu}$ is a standard extensive game with perfect
recall;

\item[(b)] if $\nu$ gives positive probability to all games in $\G$,
then $\vec{\sigma}$ is a Nash equilibrium of $\Gamma^\nu$ iff
$\vec{\sigma}'$ is a generalized Nash equilibrium of $\Gamma^*$,
where $\sigma_{i,\Gamma'}(\<\Gamma^h\>\cdot
h')=\sigma'_{i,\Gamma'}(\Gamma^h,h')$.
\end{itemize}
\eothm

\prf
For part (a), suppose that $\<\Gamma'\>\cdot h'_1)$ and
$\<\Gamma''\>\cdot h''_1)$ are in the same
$(i,\Gamma^+)$-information set
of $\Gamma^{\nu}$
and that $h'_2$ is a prefix of $h'_1$ such that
$P^{\nu}(\<\Gamma'\> h'_2)) = (i,\Gamma^+)$. By definition of
$\Gamma^{\nu}$, it must be the case that there exist $i$-information
sets $I_1$ and $I_2$ in $\Gamma^+$ such that
$\F(\Gamma',h'_1)=\F(\Gamma'',h''_1)=(\Gamma^+,I_1)$ and
$\F(\Gamma',h'_2)=(\Gamma^+,I_2)$.
If $h_1$ is a history in $I_1$, C8 implies that there
exists a prefix $h_2$ of $h_1$ such that $P^+(h_2)=i$,
$\F(\Gamma^+,h_2)=(\Gamma^+,I_2)$ and if $h'_2\cdot \<m\>$ is a
prefix of $h'_1$, then $h_2\cdot \<m\>$ is a prefix of $h_1$.
Applying C8 again, it follows that there exists a prefix $h''_2$ of
$h''_1$ such that
$P^{\Gamma''}(h''_2)=i$ and $\F(\Gamma'',h''_2)=(\Gamma^+,I_2)$ and
if $h_2\cdot \<m\>$ is a prefix of $h_1$, then $h''_2\cdot \<m\>$ is
a prefix of $h''_1$. Therefore, by definition of $\Gamma^{\nu}$,
$(\Gamma'',h''_2)$ and $(\Gamma',h'_2)$ are in the same information
set.

Suppose further that $h'_2\cdot \<m\>$ is a prefix of $h'_1$. Thus,
$h_2\cdot \<m\>$ is a prefix of $h_1$, which implies that
$h''_2\cdot \<m\>$ is a prefix of $h''_1$. This proves part (a).

For part (b), let $Pr^{\nu}_{\vec{\sigma}}$ be the probability
distribution over the runs in $\Gamma^{\nu}$ induced by the strategy
profile $\vec{\sigma}$ and $f_c^{\nu}$.
$Pr^{\nu}_{\vec{\sigma}}(z)$ is the product of the probability of
each of the moves in $z$. (It is easy to define this formally by
induction on the length of $z$; we omit details here.) Similarly, let
$Pr^{h}_{\vec{\sigma}'}$ be the probability distribution over the
runs in $\Gamma^h\in \G$ induced by the generalized strategy profile
$\vec{\sigma}'$ and $f_c^h$.
Note that if
$Pr^{h}_{\vec{\sigma}'}(z)>0$, then $z\in \lfloor \H^h \rfloor$.
Thus, $\<\Gamma^h\>\cdot z \in \H^{\nu}$.

For all strategy profiles $\sigma$ and generalized strategy profiles
$\sigma'$, if
$\sigma'_{i,\Gamma'}(\Gamma^h,h')=\sigma_{i,\Gamma'}(\<\Gamma^h\>\cdot
h')$, then it is easy to see that for all $z\in Z^h$ such that
$Pr^{h}_{\vec{\sigma}'}(z)>0$, we have that
$Pr^{\nu}_{\vec{\sigma}}(\<\Gamma^h\>\cdot
z)=\nu(\Gamma^h)Pr^{h}_{\vec{\sigma}'}(z)$. And since $\nu$ is a
probability measure such that $\nu(\Gamma^h)>0$ for all $\Gamma^h\in
\G$, we have that $Pr^{\nu}_{\vec{\sigma}}(\<\Gamma^h\>\cdot z)>0$
iff $Pr^{h}_{\vec{\sigma}'}(z)>0$. Suppose that $\vec{\sigma}$ is a
Nash equilibrium of $\Gamma^{\nu}$. Suppose, by way of
contradiction, that $\vec{\sigma}'$ such that
$\sigma'_{i,\Gamma'}(\Gamma^h,h')=\sigma_{i,\Gamma'}(\<\Gamma^h\>\cdot
h')$ is not a generalized Nash equilibrium of $\Gamma^*$. Thus,
there exists a player $i$, a game $\Gamma^+\in\G_i$, and a local
strategy $s'$ for player $i$ in $\Gamma^+$ such that

\begin{equation}\label{eq1}
\sum_{z\in Z^+}Pr^+_{\vec{\sigma}'}(z)u_i^+(z)< \sum_{z\in
Z^+}Pr^+_{(\vec{\sigma}'_{-(i,\Gamma^+)},s')}( z)u_i^+(z).
\end{equation}

Define $s$ to be a strategy for player $(i,\Gamma^+)$ in
$\Gamma^{\nu}$ such that $s(\<\Gamma^h\>\cdot h')=s'(\Gamma^h,h')$.
Multiplying (\ref{eq1}) by $\nu(\Gamma^+)$ and using the observation
in the previous paragraph, it follows that

\begin{equation}\label{eq2}
\sum_{z\in \lfloor
Z^+\rfloor}Pr^{\nu}_{\vec{\sigma}}(\<\Gamma^+\>\cdot z)u_i^{+}(z)<
\sum_{z\in \lfloor
 Z^+\rfloor}Pr^{\nu}_{(\vec{\sigma}_{-(i,\Gamma^+)},s)}(\<\Gamma^+\>\cdot
z)u_i^{+}(z).
\end{equation}

By definition of $u_{i,\Gamma'}^{\nu}$, (\ref{eq2}) holds iff

\begin{equation}\label{eq3}
\sum_{z^{\nu}\in Z^{\nu}}Pr^{\nu}_{\vec{\sigma}}(
z^{\nu})u_{i,\Gamma^+}^{\nu}(z^{\nu})< \sum_{z^{\nu}\in
Z^{\nu}}Pr^{\nu}_{(\vec{\sigma}_{-(i,\Gamma^+)},s)}(
z^{\nu})u_{i,\Gamma^+}^{\nu}(z^{\nu}).
\end{equation}

Therefore, $\vec{\sigma}$ is not a Nash equilibrium of $\Gamma^\nu$,
a contradiction. The proof of the converse is similar; we leave
details to the reader. \eprf

\ocor{cor:Nashex} Every game with awareness has a generalized Nash
equilibrium. \eocor

\prf For games with perfect recall,
there is a natural isomorphism between mixed strategies and behavioral
strategies, so a Nash
equilibrium in behavior strategies can be viewed as a Nash equilibrium in
mixed strategies \cite{OR94}.
Moreover, mixed-strategy Nash equilibria of an extensive-form game are
the same as the mixed-strategy Nash equilibria of its normal-form
representation.
Salonen \citeyear{Sal02} showed that there exists a Nash equilibrium
in mixed strategies in a normal form games
with an arbitrary set $N$ of players if,
for each player $i$,
the set $S_i$ of pure strategies of player $i$ is a compact metric
space, and the utility functions  $u_i: S \rightarrow \IR$ are
continuous for all $i \in N$, where $\IR$ is the set of real numbers and
$S = \Pi_{i \in N} S_i$, the set of pure strategies, is endowed with the
product topology.
Since in $\Gamma^{\nu}$, every
player has a finite number of pure strategies,
$S_i$ is clearly a compact metric space. Moreover,
since each player's utility depends only on the strategies of a finite
number of other players, it is easy to see that $u_i : S \rightarrow
\IR$ is continuous for each player $i \in N$.
It follows that there exists a Nash
equilibrium of $\Gamma^{\nu}$. Thus, the corollary is immediate from
Theorem~\ref{THM:NASHEX}. \eprf

\end{document}